%% file: main.tex
\documentclass[conference,10pt,a4paper]{IEEEtran}

\usepackage{url,comment,cite, footmisc}
\usepackage{amsmath,amssymb,amsfonts,dsfont}
\usepackage{graphicx,color}
\usepackage{pifont}
\usepackage{amsmath,mathtools}
\usepackage{epstopdf}

\includecomment{journal}
\excludecomment{conference}

\newcommand{\wfd}{Wi-Fi~Direct\ }
\newcommand{\wf}{Wi-Fi\ }

\begin{document}

\title{Content-centric Routing in \\Wi-Fi Direct Multi-group Networks}


\author{\IEEEauthorblockN{C. Casetti, C.-F. Chiasserini, L. Curto Pelle,
C. Del Valle, Y. Duan, P. Giaccone}
\IEEEauthorblockA{Department of Electronics and Telecommunications\\
Politecnico di Torino, Italy \vspace*{12pt}
}
}

\maketitle
\thispagestyle{empty}\pagestyle{plain}

\input{abstract.tex}

\input{intro.tex}

\input{wifi.tex}

\input{multigroup.tex}

\input{discovery.tex}

\input{performance.tex}

\input{relatedwork.tex}

\input{conclusion.tex}


\bibliographystyle{IEEEtran}
\bibliography{biblio}

%
%
%
%
\end{document}

%% file: abstract.tex
\begin{abstract}

The added value of Device-to-Device (D2D) communication amounts to an efficient
content discovery mechanism that enables users to steer their requests toward 
the node most likely to satisfy them. In this paper, we address the implementation
of content-centric routing in a D2D architecture for Android devices 
based on WiFi~Direct, a protocol recently standardised by the Wi-Fi Alliance. 
After discussing the creation of multiple D2D groups, we introduce novel 
paradigms featuring intra- and inter-group bidirectional communication.
We then present the primitives involved in content 
advertising and requesting among members of the multi-group network. 
Finally, we evaluate the performance
of our architecture in a real testbed involving Android devices 
in different group configurations. {We also compare the results against the ones achievable exploiting Bluetooth technologies.}  

\end{abstract}

%% file: intro.tex
\section{Introduction\label{sec:intro}} 


It can be argued that the vast majority of wireless communicating
devices in use today rely on an Access Point (AP)-based paradigm.
Cellular networks, Wi-Fi hotspots, all require user devices to ``associate''
to a common base station before they can operate. Undeniably, such 
paradigm is convenient: it facilitates a uniform service provision, 
it simplifies management and it is essential in case billing is required.
At the same time, it creates a cumbersome overhead for communications
which, by virtue of the location of endpoints, might best be served by a 
direct link. Exploitation of Device-to-Device (D2D) connectivity, whether 
in an unrestrained or in a network-controlled fashion, is at the forefront
of standardisation and research efforts. Such interest is spurred by
the commercial appeal and widespread availability of 
Bluetooth Low Energy~\cite{BTle} and Wi-Fi Direct~\cite{wfd}, 
technologies that smartphone and tablet manufactures are increasingly 
incorporating in their products. 

While in the past D2D communication was largely relegated to cable-replacement
use cases, today it is touted as a game-changing factor in mass communication, 
thanks to its enhanced spectral efficiency and traffic offloading capabilities. 
Some commonly envisioned scenarios for D2D are: machine-to-machine communication,
Internet of Things architectures,
infrastructure replacement (in case of failure), social content 
sharing. Additionally, D2D (and Wi-Fi Direct in particular) has the potentiality
to play a crucial role in future LTE offloading strategies. LTE standardisation
is looking at the interoperability with other D2D technologies by introducing 
the concept of network-assisted D2D communication: the cellular interface would
jump-start the D2D link between suitable devices by handling the discovery and 
authentication phases, thus serving as broker party \cite{3GPP1,Mancuso,Sergey1}.

However, many of the promises in store for D2D communication lay bare what is 
arguably its biggest flaw: lacking a ``static'' infrastructure, the availability 
of content is, at best, spotty and unreliable. Even if requested content is cached 
by a nearby node, reachable through a multi-hop D2D path, a robust content discovery
and retrieval mechanism is needed. Such mechanism should be aware, and, if 
possible, should leverage the peculiarities 
of the D2D environment: high node churn, volatile topologies and resource-constrained
devices. 

In this paper, we focus on the potentiality of Wi-Fi Direct as D2D communication
technology in medium and large-scale scenarios, using open-source, 
non-rooted Android devices. Our contribution is 
manyfold. 
\begin{itemize}
\item For starters, 
we investigate in depth the limitations that 
the current Android OS exhibits in some crucial Wi-Fi
direct features, and in the roles that devices can play in a D2D multi-device 
topology. 
\item Secondly, we work around the above limitations by designing a multi-group, interconnected
logical topology that overcomes the limitations of the physical one by exploiting
transport-layer tunneling. Such logical topology allows us to enable
bidirectional, inter-group data transfers, which would otherwise be 
impossible in today's Wi-Fi Direct-based networks. 
\item Thirdly, 
in order to address the content availability issue,
we implement a content-centric routing architecture on our D2D topology. 
In content-centric routing, 
users do not need to know the physical whereabouts of data
(as in traditional IP routing, in which the hosting device is pinpointed by
a univocal identifier), but they just focus on the content they need and let
the network do the rest. Routing tables thus carry
content-oriented routing information that reflects both (i) 
the availability of specific
content either in the local group of devices or in a nearby, reachable
group, and  (ii) the above
transport-layer tunneling mechanism through which content can be reached.
\item At last, we implement a novel content registration/advertisement protocol 
that is designed to populate Content Routing Tables (CRT) consistently with 
the data that each user is willing to share (and thus advertises in the D2D
network).
\end{itemize}
To our knowledge, our work is the first that tackles bidirectional,
inter-group communication in Wi-Fi Direct networks, and proposes and
implements a solution to support this data transfer paradigm.
Furthermore, we realised a small-scale testbed using off-the-shelf Android devices to test 
both the feasibility of our multi-group topologies, as well as the efficiency 
of content-centric routing along with the registration/advertisement protocol.

The rest of the paper is organised as follows. Section~\ref{sec:wifidirect}
provides an overview of Wi-Fi Direct. Section~\ref{sec:mechanism} highlights
some of the limitations that topology formation suffers from in WiF-Direct
devices and details the multi-group communication mechanism.
Our content-centric routing architecture and registration/advertisement protocol 
\begin{journal}
are presented in Section~\ref{sec:discovery}. 
\end{journal}
\begin{conference}
{\color{red}
are introduced in Section~\ref{sec:discovery}. 
For the sake of space, details have been omitted and reported in a companion technical report~\cite{wifi-tr}.
}
\end{conference}
Section~\ref{sec:performance}
illustrates the results derived from our testbed implementation. 
Related work is discussed in Section ~\ref{sec:relatedwork}, while
Section~\ref{sec:conclusions}
draws some conclusions and points out directions for future research.

%% file: wifi.tex
\section{The Wi-Fi Direct Technology\label{sec:wifidirect}}

\begin{figure}[!b]\centering
\includegraphics[width=6cm]{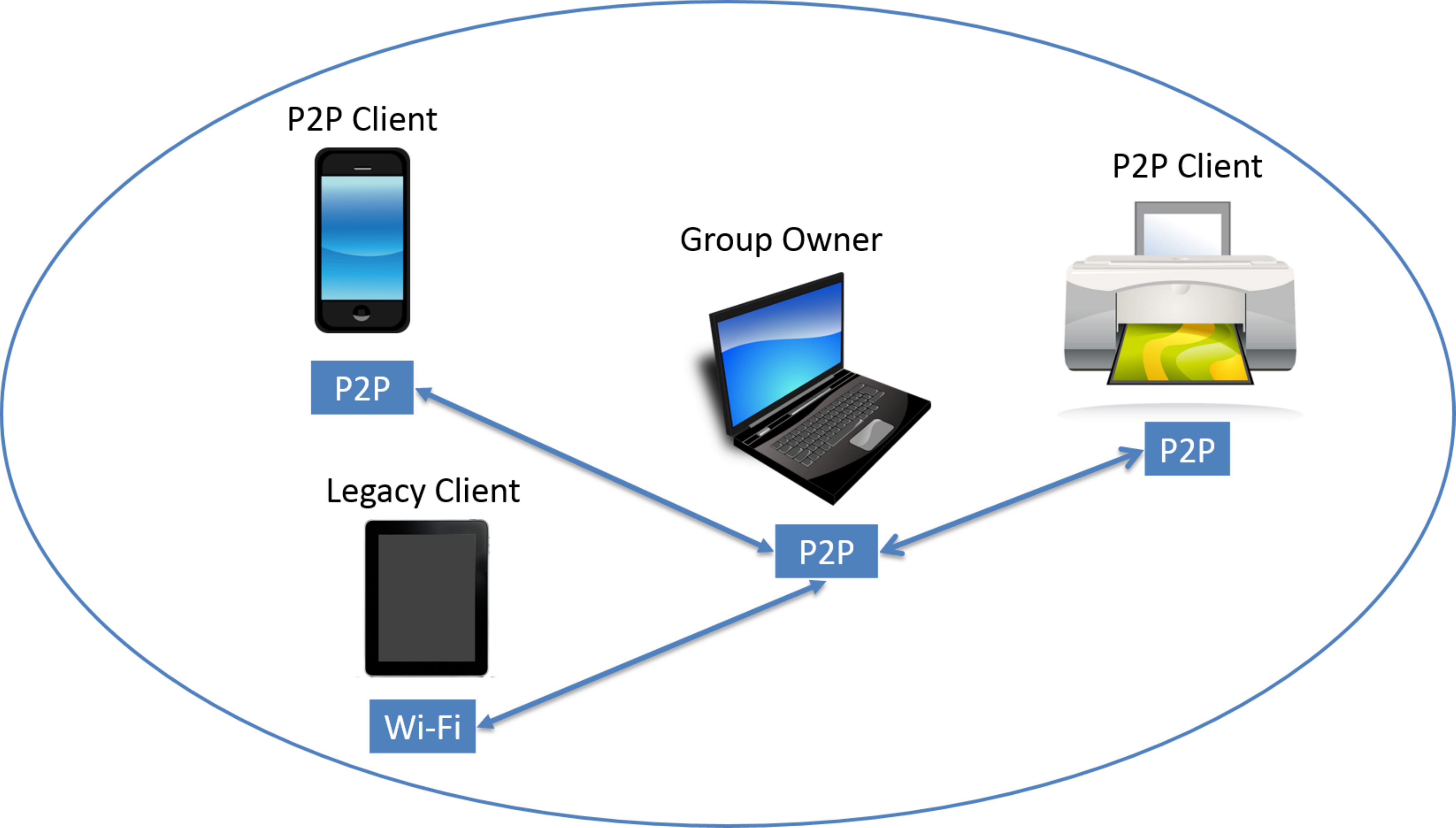}
\caption{Basic \wfd group with one GO, two P2P-clients and one legacy-client. \label{fig:Group}}
\end{figure}

\wfd is a recent protocol standardized by the  Wi-Fi Alliance~\cite{wfd}, with
the aim to
enable D2D communications between nodes, referred to as {\em peers}.
Communication among peers in \wfd occurs within a single {\em group}.
One peer in the group acts as Group Owner (GO) and  the other
devices, called {\em clients}, associate to the GO (see, e.g., Fig.~\ref{fig:Group}). 
Such roles within the group are not predefined,
but are negotiated upon group  formation. After the GO is elected,
the role of each peer remains unchanged during the whole group session. Only when
the GO leaves the group, the peers become disconnected and a new
group must be created. 

The group works as an infrastructure \wf BSS operating on a single
channel, through which the peers communicate. The GO periodically transmits
a beacon to advertise the group so as to enable other
devices to discover and, possibly, join the group. As depicted in
Fig.~\ref{fig:Group}, 
each client is either
a {\em P2P client} or a {\em legacy client}. A P2P client supports 
the \wfd protocol, whereas a 
legacy-client is a conventional \wf node that does not support \wfd
and ``sees'' the GO as a traditional \wf AP. P2P clients and
legacy clients coexist seamlessly in the same group.




It is important to note that 
\wfd has been designed to support D2D communication within a group,
however 
its protocol does not prevent the communication between different
groups. Indeed, a peer can act as a bridge between two groups, or
between the group and other networks.


\begin{figure}[!tb]\centering
\includegraphics[width=0.3\textwidth]{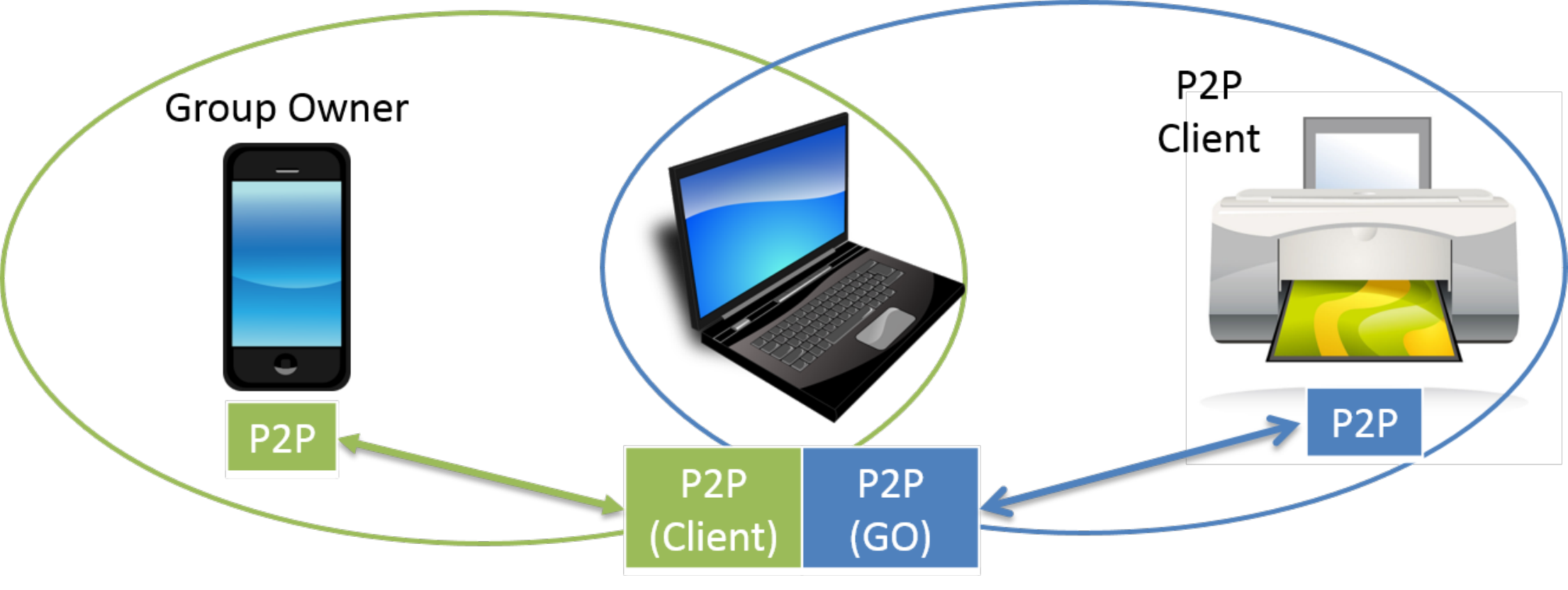}
\caption{Communication between two \wfd groups. \label{fig:g2g}}
\end{figure}

\begin{figure}[!tb]\centering
\includegraphics[width=0.3\textwidth]{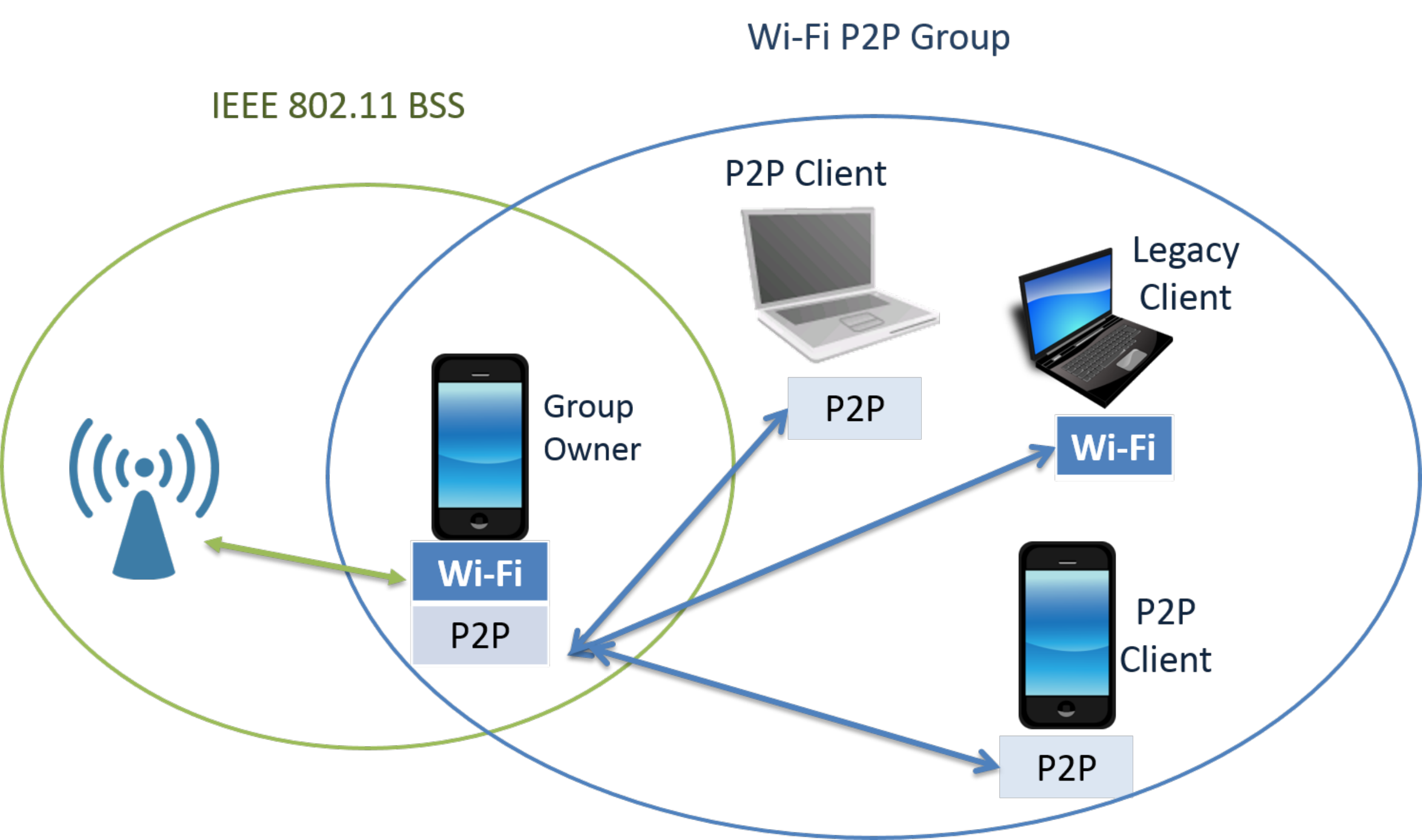}
\caption{Communication between a \wfd group and a \wf BSS. \label{fig:w2p}}
\end{figure}

One possible scenario, as shown in Fig.~\ref{fig:g2g}, consists of a
bridge peer (the middle node) behaving as GO for one group and as P2P-client in another
group. We stress that the bridge peer must support two different MAC entities
at layer 2, with two different MAC addresses. 
A peer can also act as a bridge between a \wfd group and a standard
infrastructure BSS. This concurrent operation is
shown in Fig.~\ref{fig:w2p}. Also in this case, the support for multiple
MAC entities is required.

%% file: multigroup.tex
\section{Multi-group Communication with Android devices\label{sec:mechanism}}{\color{red} }

As mentioned, we focus on user devices running an open-source Android OS 
due to their wide popularity, 
and we investigate how to provide bidirectional multi-group communication 
in networks composed of such devices. 

Android devices offer a limited, controlled set of networking capabilities
for security reasons. 
It is of course possible to ``root'' a device in order to access
advanced capabilities, but we do not take this possibility into account since
the rooting process requires skills that are beyond the average user,
and it renders the warranty null and void.
Thus, we  only act upon application-layer functionalities, i.e., no
changes can be performed at the transport or network layer (like
changing IP addresses for P2P interfaces, configuring routing tables, etc).

A multi-group topology could be implemented by letting a device have two
 virtual P2P network interfaces: in this way, it could act as a bridge 
using a different MAC entity in each group. In non-rooted Android devices,  however, the programmer
cannot create a custom virtual network interface. Our
experiments revealed that none of the following scenarios are feasible in Android,
much though they are not expressly forbidden by the standard:
\begin{enumerate}
\item a device plays the role of P2P client in one group and GO in another,
\item a device behaves as the GO of two or more groups,
\item a device behaves as client in two or more groups.
\end{enumerate}

\begin{figure}[!tb]\centering
\setlength{\unitlength}{0.01\textwidth}
\includegraphics[width=8cm]{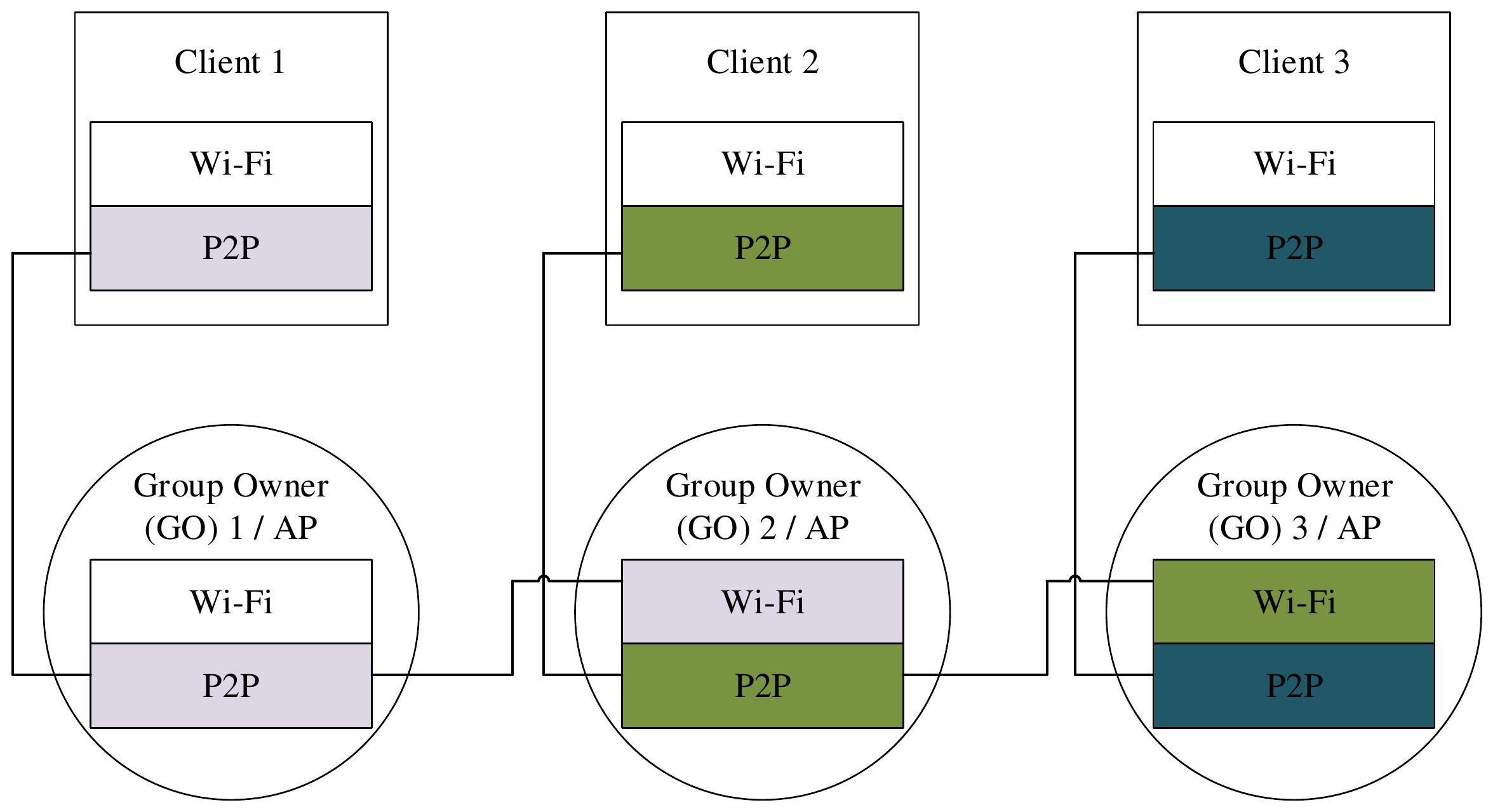}
\caption{Multi-group physical topology with six devices (three clients and three GOs). 
GO2 and GO3 are bridge nodes, i.e., they are legacy clients of GO1 and
GO2, respectively.\label{figTopophy}}
\end{figure}

Thus, in order to create a multi-group physical topology (i.e., bridge nodes), we let a GO
be a legacy client in another group. Specifically, we proceed as
depicted in Fig.~\ref{figTopophy}, where three inter-connected groups
are formed with six devices. GOs are represented by circles and clients by
squares. 
In each peer, we enable  two network interfaces, one of which is the
conventional Wi-Fi interface and the other (P2P) is used for \wfd
connection.  The interfaces used to form a group are highlighted using
the same color, while connections are represented by lines. It is
important to remark that each group represents a different Wi-Fi Basic
Service Set (BSS). Furthermore, note that GO2 and GO3 also act as legacy
clients of GO1 and GO2, respectively. GO1 is not acting as a
legacy client since it is not associated to any other group. As
discussed later in Section~\ref{sec_log}, the fact that one GO is a legacy client
of another GO affects its forwarding capabilities.


For ease of presentation, in the following we often take the three-group
physical topology depicted in Fig.~\ref{figTopophy} as reference
scenario and refer to the three groups as Group 1, Group 2 and Group 3,
respectively.


\subsection{IP address assignment\label{sec_ip}} 

In Android devices, once a \wfd connection is established, the
GO  automatically runs the DHCP to assign IP addresses to
itself (192.168.49.1/24) as well as to the P2P clients or legacy
clients in its own group
(192.168.49.$x$/24 where $x$ is a random  
number $\in [2,254]$ to minimize the chance of address conflicts). 
Therefore, the P2P interfaces of all 
GOs have the same IP address, namely 192.168.49.1. The Wi-Fi
interfaces of the GOs that act as legacy clients in another
group are assigned an IP address in the format 192.168.49.$x$/24. 
Similarly,  P2P interfaces of clients are assigned 
different IP addresses in the format 192.168.49.$x$/24.

\begin{figure}[!tb]\centering
\setlength{\unitlength}{0.01\textwidth}
\includegraphics[width=8cm]{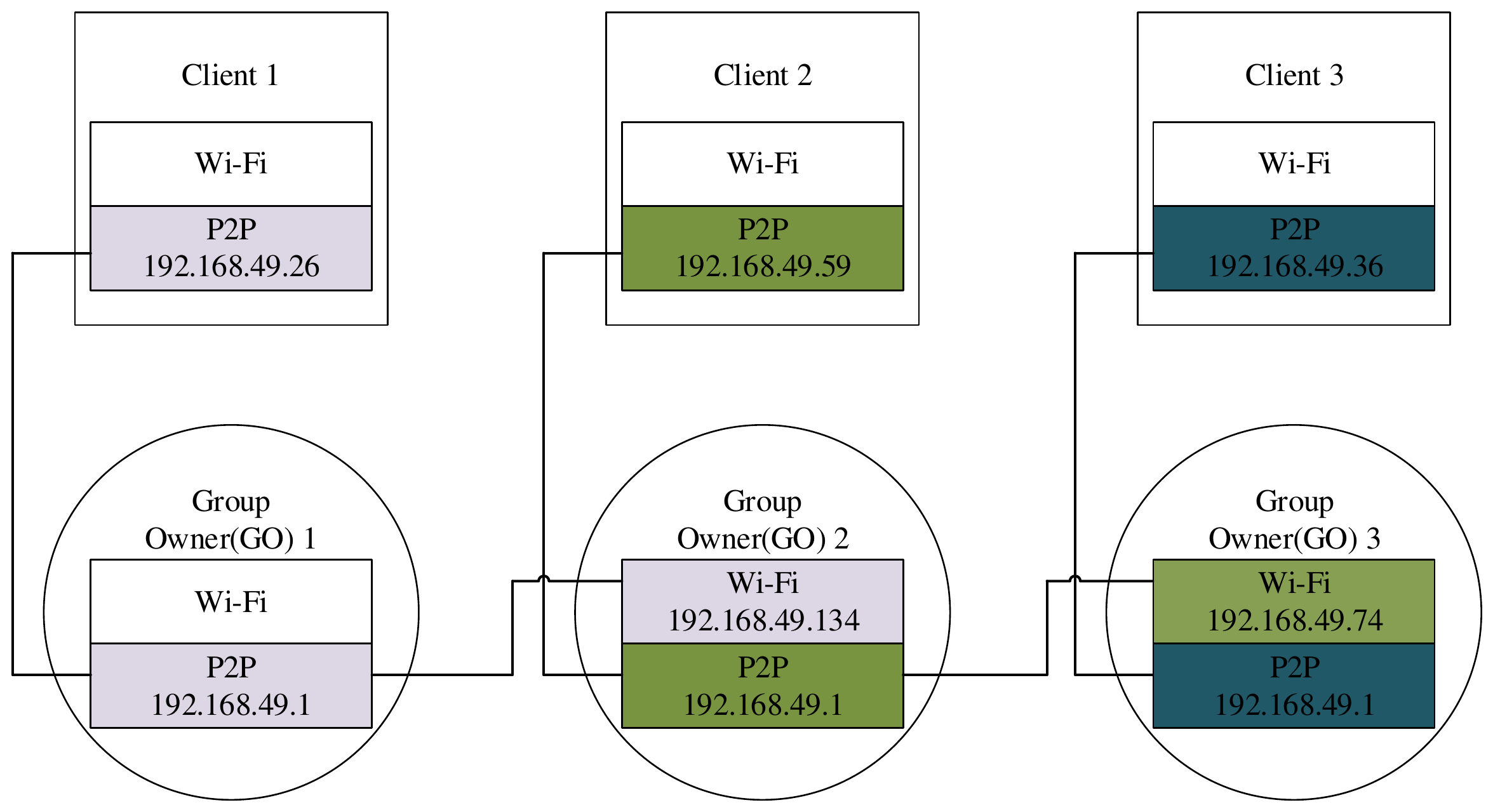}
\caption{Example of IP addresses (/24) for multi-group configuration.\label{figTopoip}}
\end{figure}

An example of IP assignment for the three-group topology is shown in Fig.~\ref{figTopoip}, 
which highlights the address conflicts for the P2P interface of the GOs.
Since GO1 is not associated with 
a \wf AP, no IP address is assigned to its Wi-Fi interface. 

\subsection{Design of the logical topology\label{sec_log}}

Given the above assignment of IP addresses, we show how to design a {\em
logical topology} that  implements multi-group communication. Our
methodology overcomes the limitations of the physical topology and of its
addressing plan, which prevent data transfers on some D2D links.

Let us start by discussing intra-group communication, as it is the basis
for enabling bidirectional inter-group communication. Two cases have to
be distinguished.

\begin{figure}[!tb]\centering
\setlength{\unitlength}{0.01\textwidth}
\includegraphics[trim=1cm 0.5cm 1cm 0.3cm, scale=0.5]{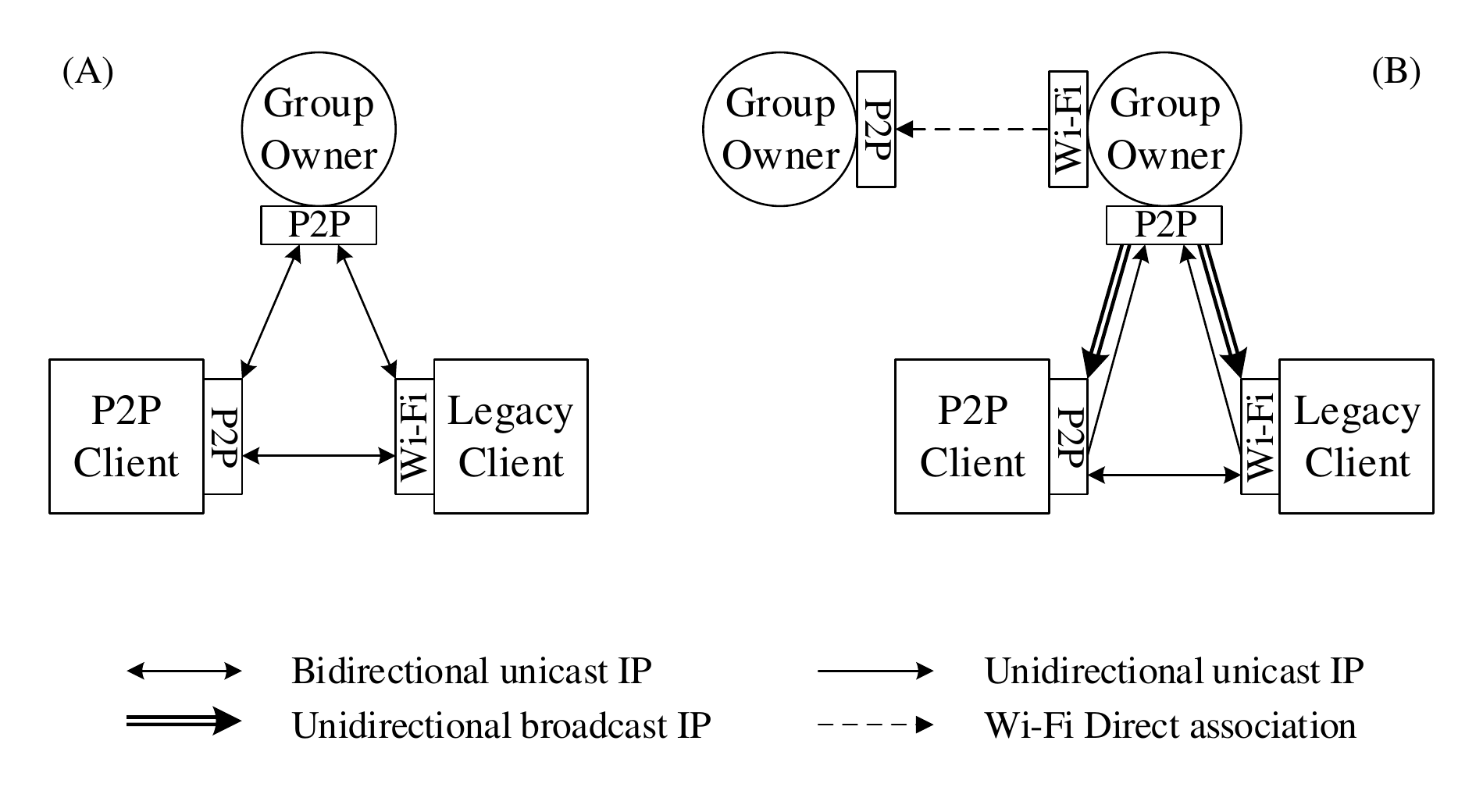}
\caption{D2D intra-group communications: (A) in an isolated group, (B)  
in a group whose GO is a legacy client in another group.\label{fig:intra}}
\end{figure}


In the first case, depicted in Fig.~\ref{fig:intra}(A), the GO is not
connected to any other group as legacy client. 
Since \wfd has been designed to provide full
connectivity among all nodes of an isolated group, all possible D2D
communications are enabled. Thus, any pair of devices (GO,  P2P clients 
and legacy clients) can exchange data at the IP layer. Note that, in the specific example in
Fig.~\ref{figTopoip}, Group 1 falls in this case (hence all  D2D
communications are allowed) since GO2 is a standard legacy client as far as
GO1 is concerned.

In the second case, illustrated  Fig.~\ref{fig:intra}(B), the GO is also
connected to another group as a legacy client. Referring again to the
example network in Fig.~\ref{figTopoip},  Group 2 and Group 3 fall in
this case. All  D2D unicast data transfers among clients (P2P or legacy
clients) are allowed, thus TCP connections and/or UDP flows between
clients are supported. Instead, between two GOs, or between a GO and its
clients, {\em only a subset of D2D data transfers are allowed}. The
reasons underlying this limitation are two. First of all, two
neighboring GOs cannot communicate directly, because of the IP address
conflict. Note that in this case one of the GOs acts as legacy client of
the other GO, as in the example of Fig.~\ref{figTopoip} where GO2 is legacy
client of GO1.  When GO2 wishes to transmit an IP packet to GO1, the
destination is set to 192.168.49.1 and the packet is thus sent to its
local loop and not to the \wf interface. Also, when GO1 sends an IP
packet to GO2 (192.168.49.134), GO2 discards it since its IP layer
detects that the packet source address matches its own
(192.168.49.1). 
The second reason pertains to the ordering of routing table entries
in the GO, as implemented by the Android OS.
When the GO wants to send a unicast IP packet to any client of its group, the packet
is invariably sent through the GO \wf interface, since the latter entry
is listed with higher priority than the P2P interface in the routing
table of the device\footnote{We consistently observed this behavior for
different devices, of different brand, running Android 4.3 and 4.4.\label{fn:repeat}}. In
the client-to-GO direction, instead, the
communication is allowed since client routing
tables list only one interface and no conflict occurs. In summary,
{\em bidirectional unicast data transfer between GO and its clients is
not allowed}, only unidirectional unicast communication between the
client and the GO can take place. Hence, no TCP connection can be
established between the GO and its clients, whereas UDP flows are
allowed only from the clients towards their own GO. 

\begin{figure}[!tb]
\centering
\hspace*{-0.5cm}
\setlength{\unitlength}{0.01\textwidth}
\includegraphics[scale=0.5]{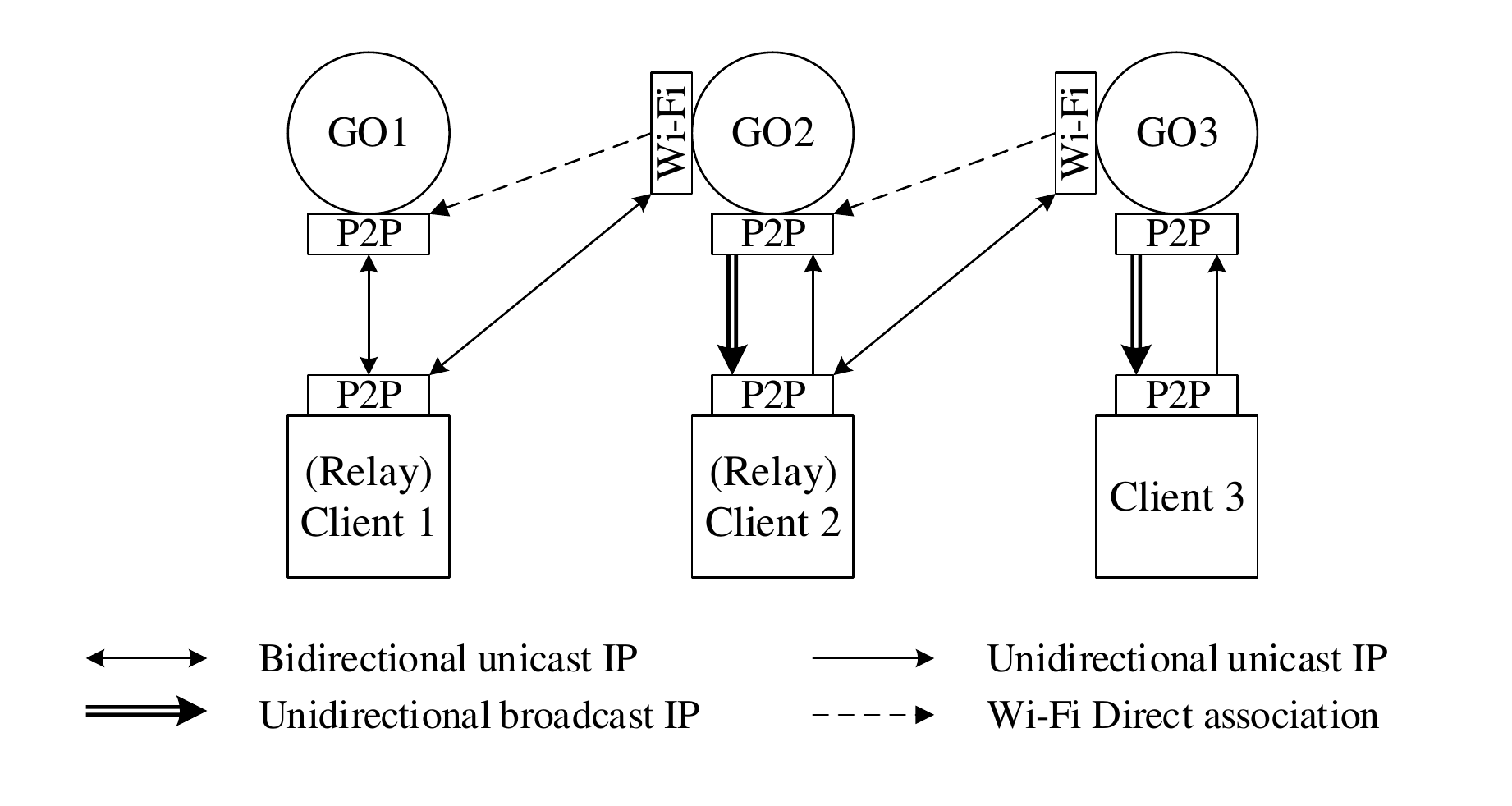}
\caption{D2D inter-group communication.
The picture refers to the example network with three groups and a linear topology. 
The P2P clients in Groups 1 and 2 are used as  relays to reach the right side group 
(Group 2 and 3, respectively). GO2 and GO3 are used as relays to reach their 
left side group (Group 1 and 2, respectively). The adopted line style follows the 
convention of Fig.~\ref{fig:intra}.\label{fig:scheme}}
\end{figure}

Conversely, broadcast IP packets sent by the GO are
always\footref{fn:repeat} sent through its P2P interface. This is an
important observation as it allows the support of bidirectional data
transfer between each client and its GO: broadcast IP packets can be
used from the GO to the clients, while unicast IP packets can be
adopted to transfer data from the clients to the GO.  Note that
broadcast packets generated by the GO will also reach the GOs associated
to it as legacy clients, but then such packets will be discarded because
of the conflict of source IP address, as discussed above. So, it is not
possible for a GO to directly reach  neighboring groups. Lines connecting
the nodes in Fig.~\ref{fig:intra}(B) summarize the possible intra-group
data transfers at IP level.

We can now focus on enabling inter-group communication in light of the
issues discussed above. We  recall that D2D communications are allowed
between any two clients within the same group (i.e., not involving the
GO at IP layer). Thus, also the communication between a P2P client and a
legacy client that is also the GO of a different group is allowed in both
directions. This observation is crucial, since it provides support for
our novel design that exploits {\em a client within the group as relay
to reach a neighboring group}. Specifically, we provide bidirectional,
inter-group communication between neighboring groups by adopting the
communication scheme shown in Fig.~\ref{fig:scheme}. To
send data from the central group (Group 2) to its right side group
(Group 3), we leverage a P2P client (Client 2) to relay the traffic
toward GO3. Instead, to send data from Group 2 to its left side group
(Group 1), GO2 itself is responsible to relay traffic toward a client in
the left side group (Client 1). 

In other words,  we build a logical topology based on transport-level
tunnels enabled by IP and MAC-layer connectivity, as follows.
\begin{itemize}
\item Unidirectional UDP tunnels between a GO and its P2P clients (e.g.,
GO1 and Client 1). They are based on broadcast IP packets from the GO to
 clients and on  unicast IP packets from  clients to the GO. When
reliable communication is required towards a single client, the GO can
adopt a classical stop-and-wait protocol.
\item Bidirectional UDP or TCP tunnels between P2P clients and legacy
clients within  the same group (e.g., between Client 1 and GO2, or
Client 2 and GO3). 
\end{itemize}

Full connectivity among nodes in a multi-group network can thus be provided
by leveraging a proper sequence of transport-layer tunnels established in the
logical topology, and switching packets at the application layer (i.e., {\em
without rooting the devices}).

\subsection{The role of the relay client}

To define a routing process that properly leverages the above
transport-layer tunnels, 
we select one client within each group,
to act as a relay node with respect to
neighboring groups. We name such node  {\em relay client}. 
In the example in Fig.~\ref{figTopoip}, Client 1
(Client 2) is the relay client connecting Group 1 (Group 2) to Group 2
(Group 3).

We implemented a basic election scheme at the  application
layer; more sophisticated solutions could be devised so as to design
smart network topologies. According to our scheme, 
the GO sends a message to one of its clients, chosen at
random among those that do not act as GO in another group, to elect it. 
To reach the desired
client, the message is sent via a broadcast IP packet through the P2P
interface. Indeed, if a unicast IP packet were used, it could 
be wrongly sent to the \wf interface (specifically, when the GO is
also legacy client in another group). 
Note that the role of each client in the group, as well as in 
other groups, is known to its GO through
\begin{journal} 
application-layer signalling (see Sec.~\ref{sec:discovery}.B).
\end{journal}
\begin{conference}
{\color{red}application-layer signalling, whose implementation details are available in~\cite{wifi-tr}.
}
\end{conference}

\begin{figure}[!tb]\centering
\includegraphics[width=8cm]{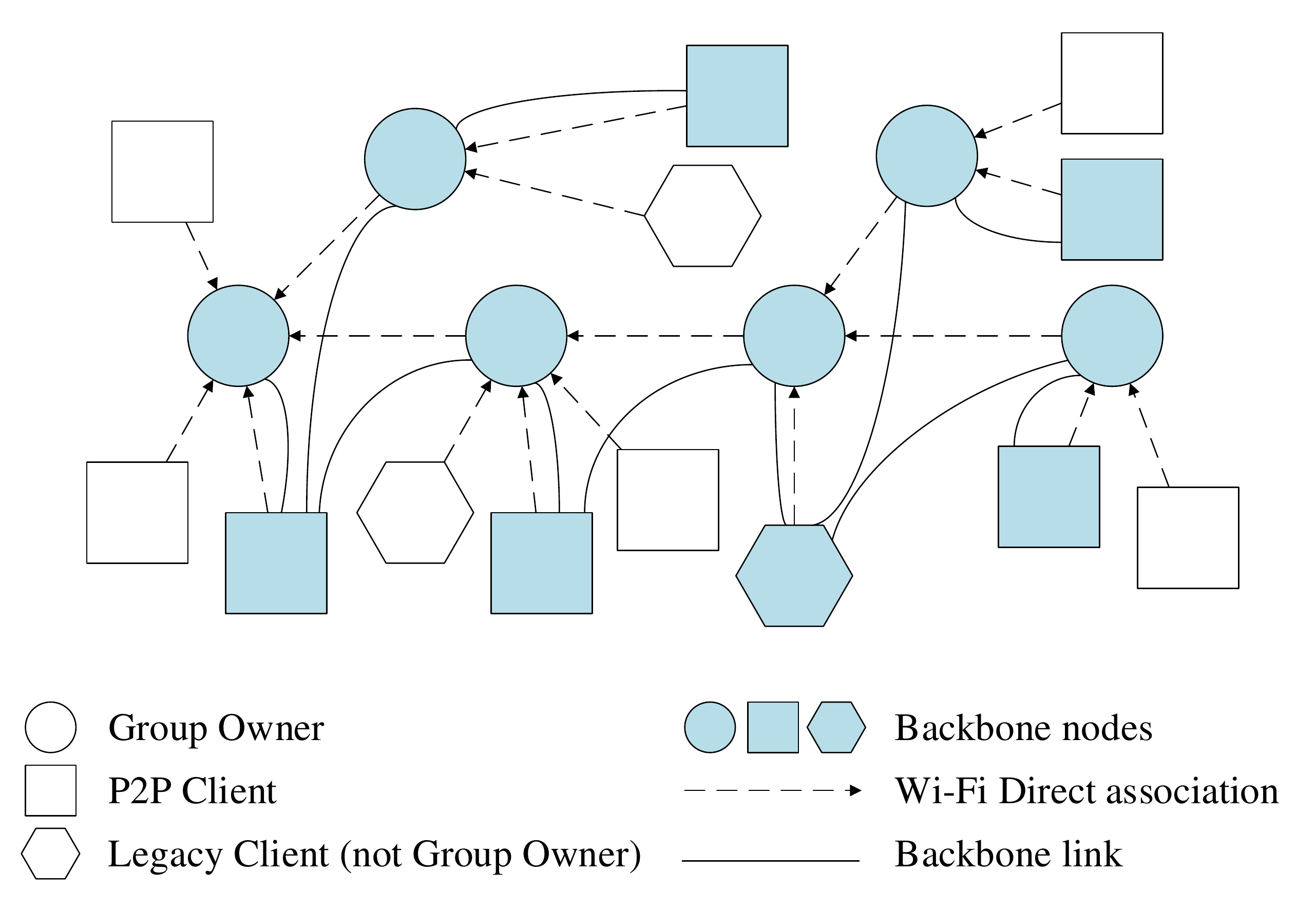}
\caption{Communication backbone over an arbitrary network topology.\label{fig:tree}}
\end{figure}

\subsection{The communication backbone\label{subsec:backbone}}

To disseminate data across a large set of devices, we then propose a
logical tree topology, connecting all groups by extending the
approach shown in Fig.~\ref{fig:scheme} to an arbitrary number of
groups. 

By doing so, we  build a communication backbone, as depicted in Fig.~\ref{fig:tree}. The figure
highlights in grey the GOs and the relay clients that compose the
backbone and provide connectivity to all other clients (P2P and \wf
clients that do not act as GOs, i.e., that are not involved in the
traffic relay process). In principle, our approach might scale
indefinitely, even if we were able to validate it  experimentally only
for few groups, as shown in Section~\ref{sec:performance}. 

It is important to remark that a path over the backbone involving
transfers from GO to  relay Client within the same group requires a
broadcast IP transmission for each of such transfers. Instead, transfers
from  relay client  to  GO do not require any  broadcast IP
transmission.

\begin{journal}

\begin{figure}[!tb]\centering
\setlength{\unitlength}{0.01\textwidth}
\includegraphics[scale=0.5]{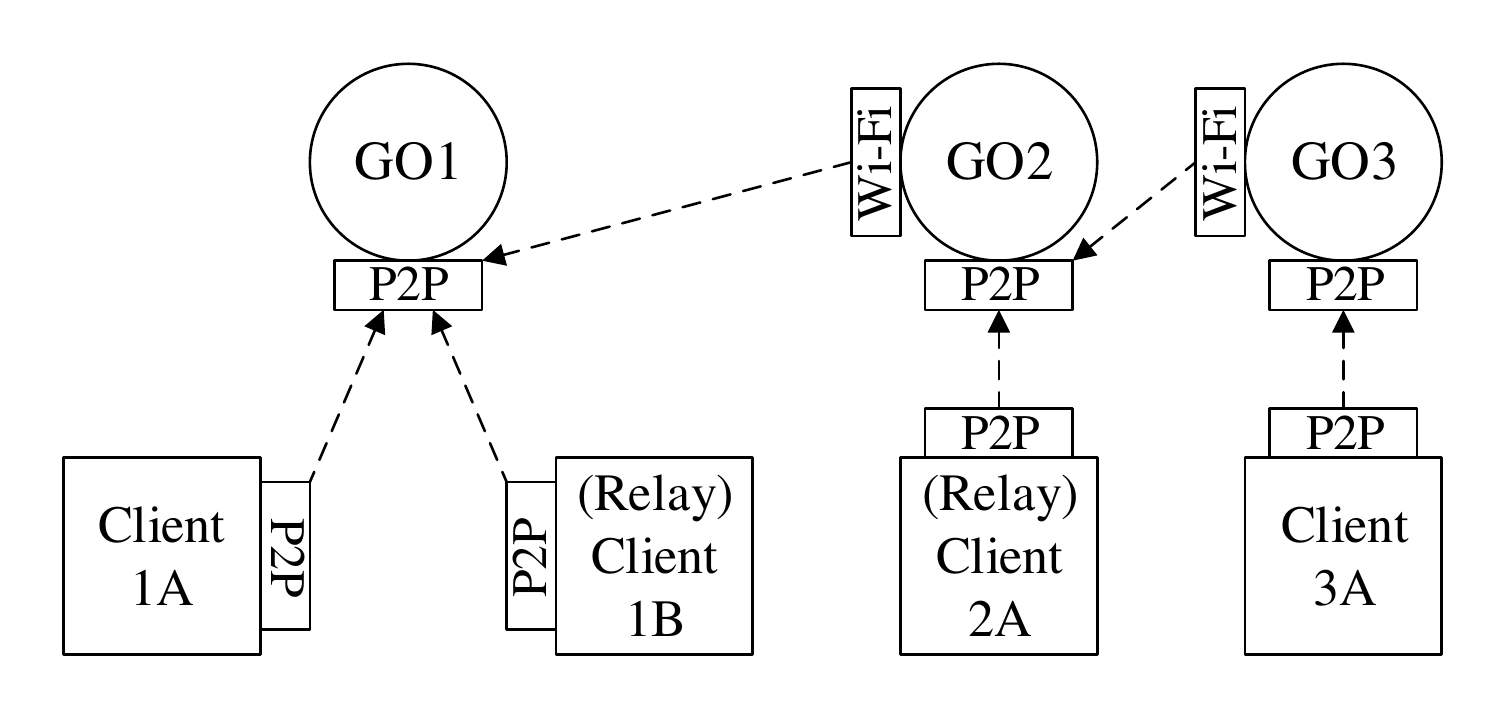}
\caption{Example scenario with 7 nodes, distributed into 3 \wfd groups. 
We only show  the associations at the \wfd level.}\label{fig:example}
\end{figure}

\begin{figure*}[!tb]\centering
\setlength{\unitlength}{0.01\textwidth}
\includegraphics[width=15cm]{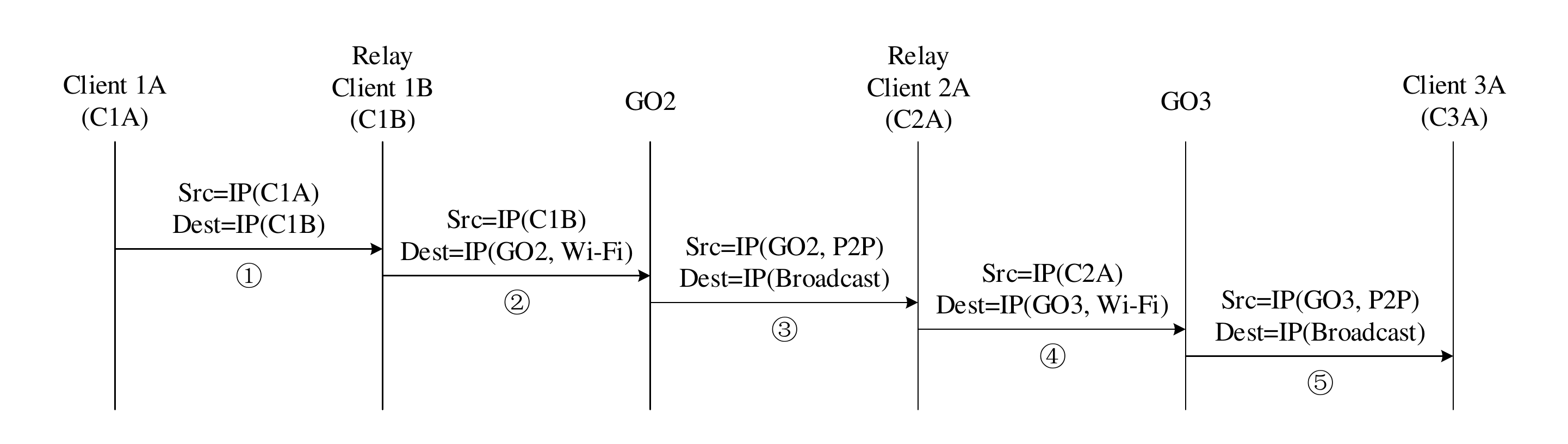}
\caption{Packet delivery at IP layer in the example network in
  Fig.~\ref{fig:example}. The packet transfer occurs from  Client 1A
  to Client 3A. 
The number is the IP packet identifier.\label{fig:test1_ip}}
\end{figure*}

\begin{figure*}[!tb]\centering
\setlength{\unitlength}{0.01\textwidth}
\includegraphics[width=15cm]{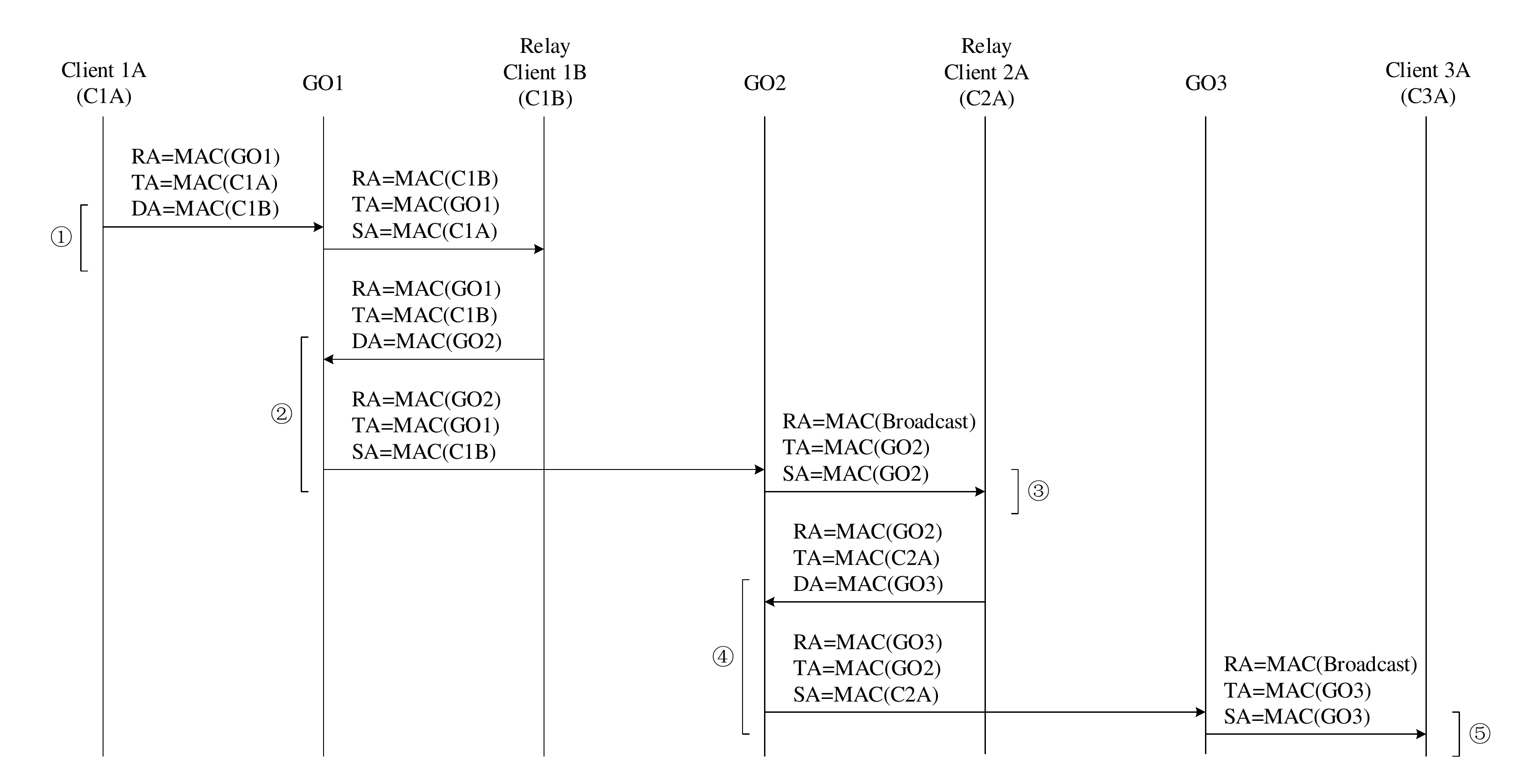}
\caption{Packet delivery at MAC layer referring to a data transfer from  Client 1A
  to Client 3A (see Fig.~\ref{fig:example}); for clarity, IP packet
  identifiers are also reported. 
\wf addresses are: TA=transmitter, RA=receiver, SA=source, DA=destination. 
\label{fig:test1_mac}}
\end{figure*}

\begin{figure*}[!tb]\centering
\setlength{\unitlength}{0.01\textwidth}
\includegraphics[width=15cm]{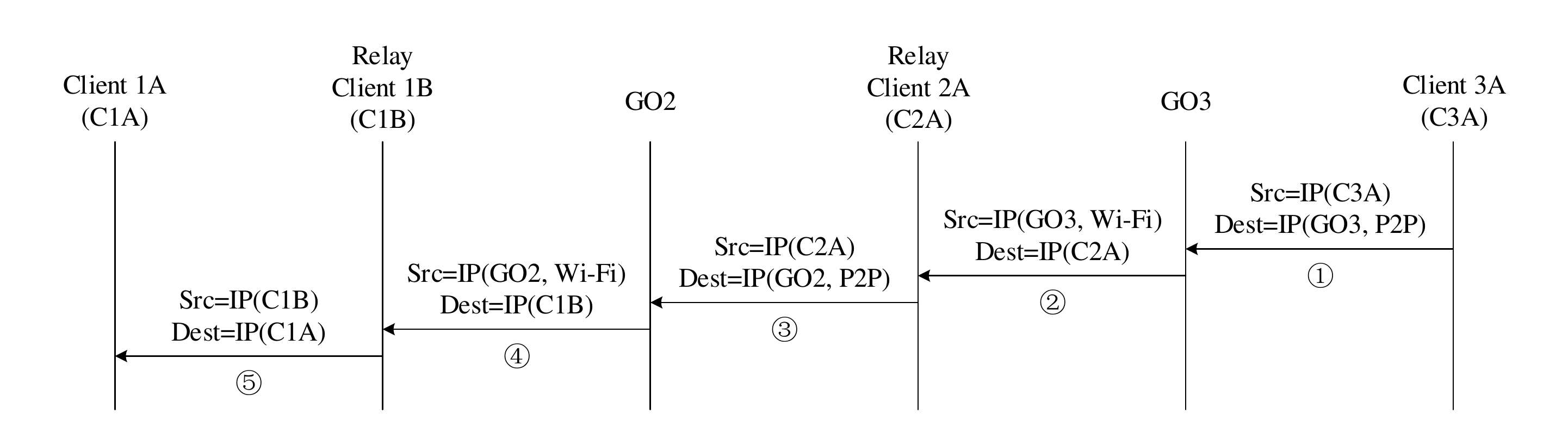}
\caption{Packet delivery at IP layer, referring to a transfer from
  Client 3A to Client 1A (see Fig.~\ref{fig:example}).  The number is
  the IP packet 
identifier.\label{fig:test2_ip}}
\end{figure*}

\begin{figure*}[!tb]\centering
\setlength{\unitlength}{0.01\textwidth}
\includegraphics[width=15cm]{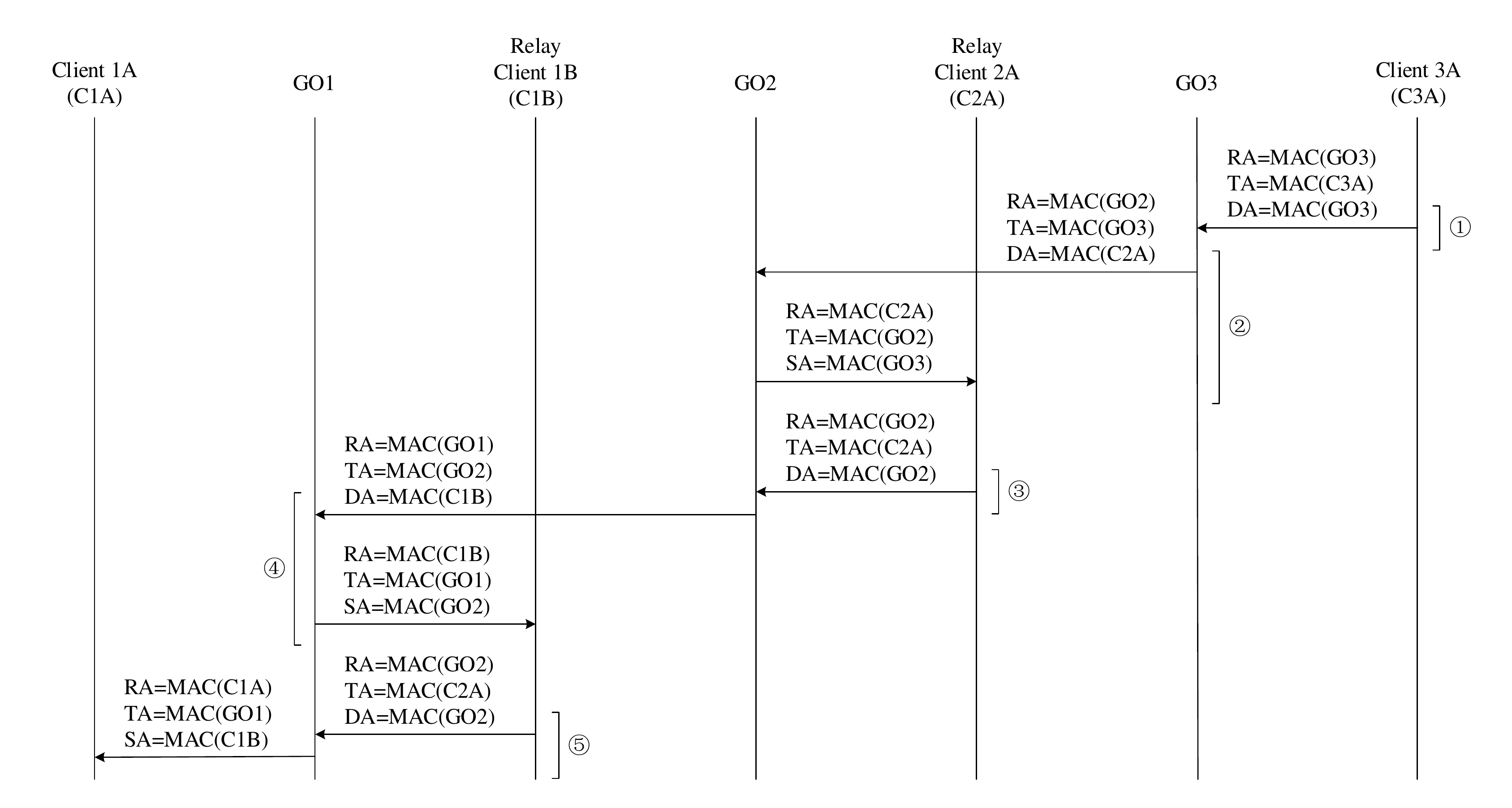}
\caption{Packet delivery at MAC layer, with IP packet identifiers. 
The diagram  refers to the data transfer from Client 3A to Client 1A.
\wf addresses are: TA=transmitter, RA=receiver, SA=source, DA=destination.\label{fig:test2_mac}}
\end{figure*}

\subsection{A step-by-step example\label{subsec:eg.service.1}}

For the sake of clarity, here we report a detailed description of the
packet forwarding process in the scenario depicted in
Fig.~\ref{fig:example}, when a packet flows from Client 1A to Client 3A,
and viceversa, using UDP as transport layer. To help
elaborate the procedure clearly, the message exchange at IP and MAC
layers are shown in Figs.~\ref{fig:test1_ip}-\ref{fig:test2_mac}. Note
that, since our focus is on content-centric routing, here we assume
devices to be aware of the next-hop to contact in order to reach a
content. The selection of the next-hop is determined by the sequence of
transport-layer tunnels composing the network backbone. For instance,
the path from Client 1A consists of the following hops: Client 1A
$\to$ Client 1B  $\to$ GO2 $\to$ Client
2A $\to$ GO3 $\to$ Client 3A. 
The next section explains the acquisition of routing information by devices.

Below, we detail the steps that let a packet be delivered from Client 1A to Client 3A.
\begin{enumerate}
\item
Client 1A encapsulates the data in the payload of a unicast UDP packet
and sends it directly  to the relay Client 1B. This packet is sent at
MAC layer to GO1, which behaves as an AP and re-sends it to the relay
client.
\item 
Client 1B processes the packet at the application layer and duplicates
the payload into a new UDP packet, sent directly to the IP of the \wf
interface of GO2. At MAC layer, the  packet is sent to  GO1, which
re-sends it to the desired GO2 interface. 
\item GO2 processes the packet at the application layer and duplicates
the payload into a new UDP packet. The UDP packet is sent as a broadcast
IP packet through the P2P interface of GO2, thus reaching the relay
client in Group 2. 
\item Client 2A processes the packet at the application layer and
duplicates the payload into a new UDP packet, to be sent directly to the
IP address of the \wf interface of GO3. 
\item Finally, GO3 processes the packet at the application layer and
duplicates the payload into a new UDP packet, sent directly to
destination Client 3A.
\end{enumerate}

In case of a packet flowing from Client 3A to Client 1A, 
the following procedure takes place. For brevity, only IP layer packet
transfers are highlighted.
\begin{enumerate}
\item Client 3A encapsulates the data in the payload of a unicast UDP
packet and sends it directly to GO3. 
\item GO3 (which is also a legacy client in Group 2) processes the packet at the application layer and duplicates
the payload into a new unicast UDP packet, sent to the relay Client 2A. 
\item In its turn, Client 2A processes the packet at the application
layer and creates a new UDP packet destined to the P2P interface of GO2. 
\item GO2 (which is also a legacy client in Group 1), again, 
creates a new UDP packet and sends it to the relay
Client 1B. 
\item Finally, Client 1B forges a new UDP packet for the final destination, Client 1A. 
\end{enumerate}
Note that, in accordance with our discussion in Section~\ref{subsec:backbone}, 
the first example above implies two IP broadcast
transmissions (i.e., transfers from a GO to a relay client within the
same group), while the second example does not involve any broadcast
IP packet. 

\end{journal}

%% file: discovery.tex
\section{Content-centric Routing}\label{sec:discovery}

We propose a network architecture in which content delivery leverages
the above forwarding scheme through a
content-centric approach. 

We assume that each node knows
the neighboring node (next hop) to which it has to send the request for
a specific content. How this knowledge is acquired is explained later in
this section. When the request reaches the node with the desired content
through a sequence of transport-layer tunnels, the content data is
forwarded back to the requester, along the same path (i.e., sequence of
tunnels) followed by the  request packet. Note that this scheme is
compatible with possible caching solutions adopted at intermediate
nodes; however,  for ease of presentation,  we will assume that each
content is provided by exactly one node in the network and that, when
such node disconnects, the content becomes unavailable.

\begin{conference}
{\color{red}
The {\em Content Routing Table} (CRT) is the data structure that provides the routing information
to reach content items. 
The information present in the CRT is updated only when new content
becomes available, or a content item becomes unavailable, following the protocol 
described  in Section~\ref{sec:adv}. Another required data structure is the {\em Pending Interest Table} (PIT),
derived from CCN~\cite{ccn}, which provides the information to route a
content to the requester.
The details on the information stored each data structure,   on the update protocol and on the adopted packet format are reported in~\cite{wifi-tr}. 
}
\end{conference}

\begin{journal}

Two main data structures are responsible for content routing, as
detailed below.

The {\em Content Routing Table} (CRT) provides the routing information
to reach content items. For each item, identified by the MD5 hash 
of its name, the CRT stores the IP address of the next-hop node to
reach the content provider, similarly to the ``gateway'' field of a
traditional IP routing table. Note that the next-hop node definition
must be tailored to the data forwarding scheme described in
Section~\ref{sec:mechanism}.

Let us examine the possible next-hop values that can be associated 
to a specific content item. To this end, we consider two neighboring
groups,  
$G$ and $\hat{G}$, and  denote by RC($G$) the relay client 
of group $G$ and by GO($G$) the GO of $G$; a similar notation 
is used for devices in $\hat{G}$. 
Let us assume that the user requesting the content is in group $G$. 
The CRT entry on the requesting user device
will associate the following possible next-hop values to the requested item.
\begin{enumerate}
\item The content item is available in $G$. Then, the next hop is 
set as the IP
address of the client in $G$ providing the content. 
\item The content item is available in $\hat{G}$, and 
$\hat{G}$ is reachable through RC($G$). In other words, GO($\hat{G}$) is a 
bridge node (i.e., it is also a legacy client in $G$), and, according to our
forwarding scheme, RC($G$) can relay traffic to it.  Thus, the next-hop for the 
requesting client, and
for all group members with the exception of GO($\hat{G}$), is RC($G$).  
The next-hop for RC($G$) is the IP address
associated with the \wf interface of GO($\hat{G}$). In the example of
Fig.~\ref{fig:example}, for a content available in Client 3A, the
next-hop for Client 1A and GO1 is the relay client 1B; the next-hop
for Client 1B is the \wf interface of GO2. 
\item The content item is available in $\hat{G}$, and
$\hat{G}$ is reachable through GO($G$), i.e., GO($G$) is a bridge node
as it acts as legacy client in  $\hat{G}$. The next-hop for the 
requesting client, and
for all members of $G$, 
is the IP address of the P2P
interface of GO($G$). The next-hop of GO($G$) is the IP
address of RC($\hat{G}$). In the example of   
Fig.~\ref{fig:example}, for a content available at Client 1A, the
next-hop for Client 3A is the P2P interface of GO3; the next-hop for GO3
is Client 2A.
\end{enumerate}
If no match is found in the CRT of an intermediate node, the packet is discarded and a
notification message is returned to the requester. 

The information present in the CRT is updated only when new content
becomes available, or a content item becomes unavailable, according to the
protocol described in Section~\ref{sec:adv}. Note that the proposed scheme
can be easily extended with a weighted list of next-hops to support multiple
copies of the same content, in case of cooperative caching
mechanisms implemented in the network. Furthermore, standard methods to
aggregate content routes in the CRT can be implemented to reduce the CRT
size and propagate differential updates. Such methods are
complementary to our scheme and out of the scope of this paper.

The second data structure is the {\em Pending Interest Table} (PIT),
derived from CCN~\cite{ccn}, which provides the information to route a
content to the requester. Before an intermediate node forwards a content
request, the node stores the IP address of the node interface from which
the request was received. Devices thus record the
``previous hop'' for each requested content item and forward the content
back to the requester upon receiving it; then, the corresponding entry
is removed from the PIT. Note that, at a given device, there may be
multiple pending requests for the same item. In this case, when the
item reaches the device, the latter forwards it toward all requesting
devices, duplicating it and sending it over the previous hops
from which the corresponding requests were received. 
A timeout is set to
remove pending requests for unavailable content. A content received by
an intermediate node without any matching entry in the PIT is discarded.
 
The flow chart in Fig.~\ref{fig:ppro} summarises the packet processing
at an intermediate node of the communication backbone (relay client or GO), when a content data packet or a content
request packet is received. 

\begin{figure}[!ht]
	\centering
	\includegraphics[width=9cm]{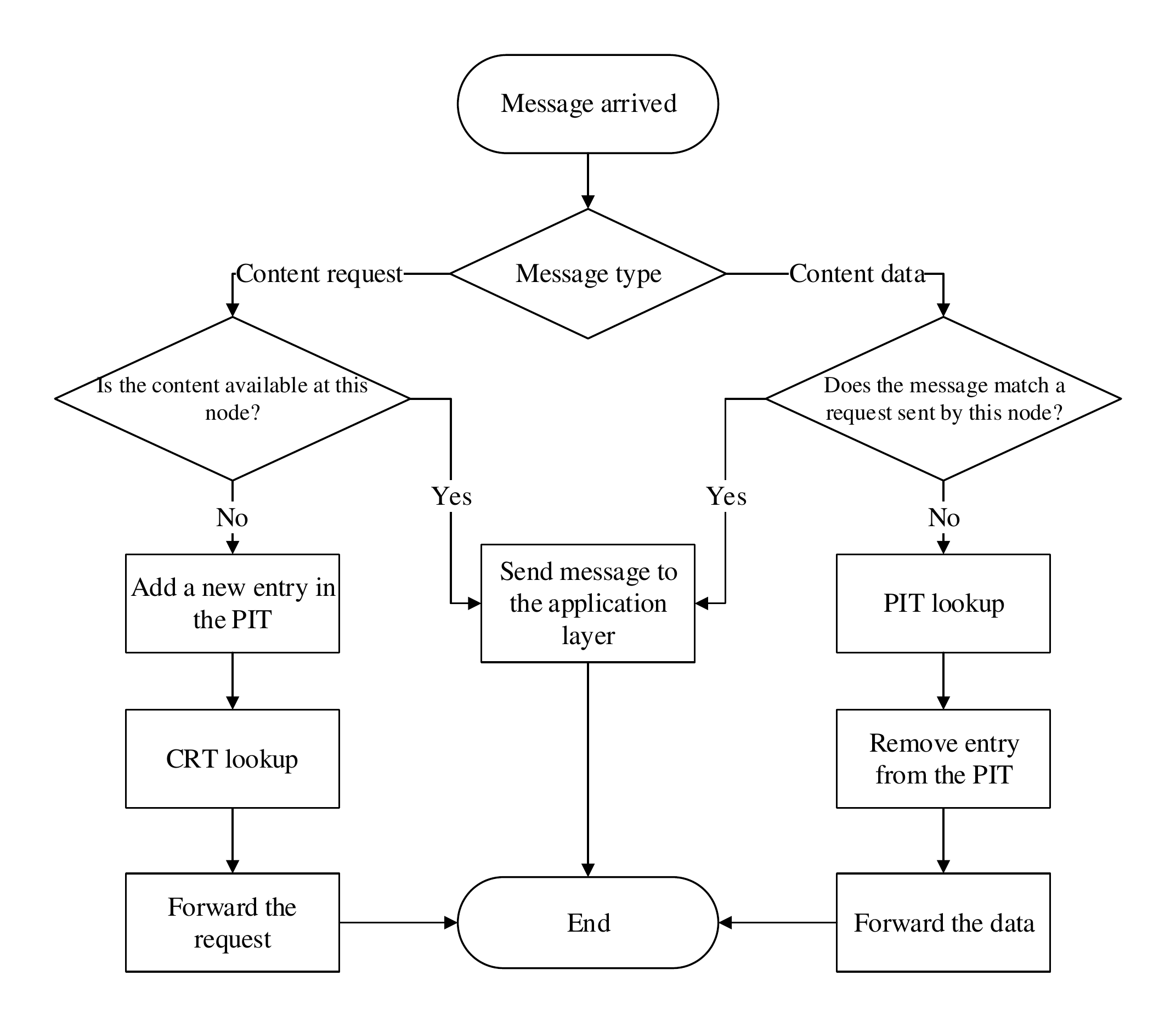}
	\caption{Packet processing at each intermediate node of the communication backbone (relay client or GO).\label{fig:ppro}}
\end{figure}

\end{journal}

\subsection{Content registration, advertisement and request}\label{sec:adv}

\begin{figure}[!ht]
	\centering
	\includegraphics[scale=0.4]{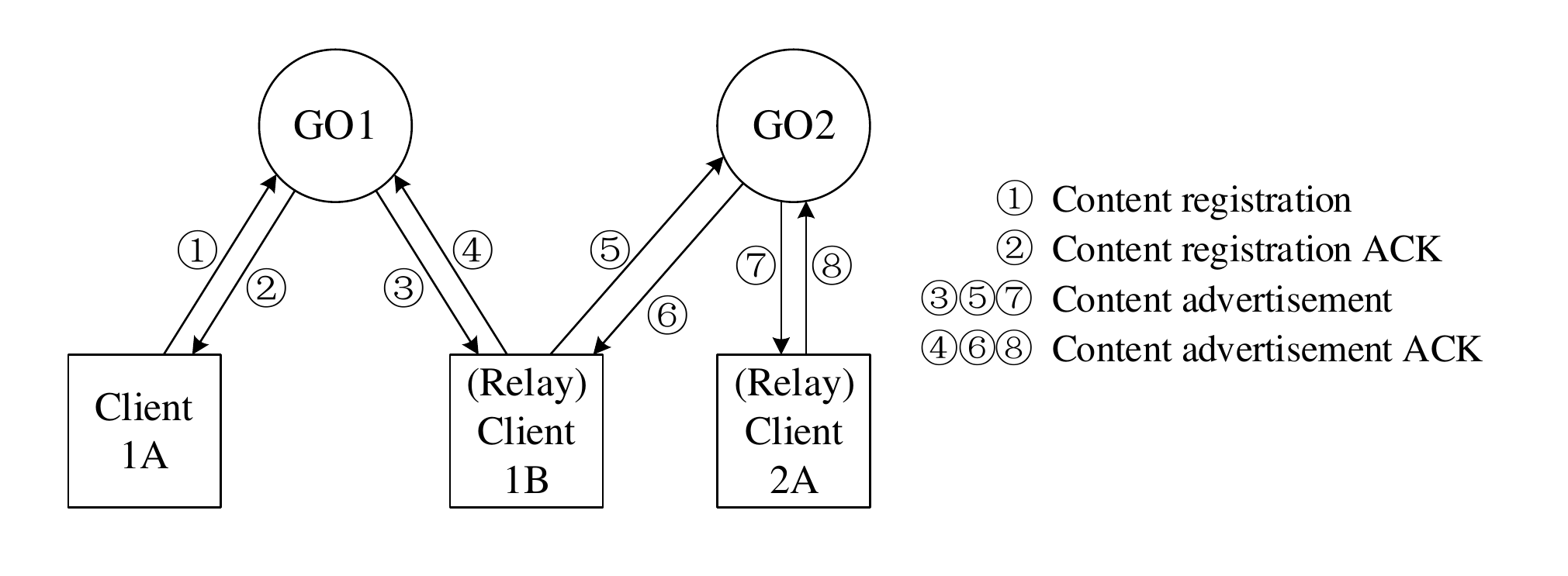}
	\caption{Message exchange triggered by a new content available in Client 1A. \label{fig:proto}}
\end{figure}

To build the CRT, we adopt a simple protocol based on the following two
phases.
{\em Content registration} is the initial phase in which a
client advertises the availability of new content within the group.
The message is sent from the client to the GO, which returns an
acknowledgment (ACK), 
guaranteeing reliable registration of content.
{\em Content advertisement} is the subsequent phase in which 
content is advertised internally and externally to the group. First, the
GO sends a broadcast message  to all (p2p and legacy) clients, to
update their CRT and waits for an ACK from the relay client. 
Thanks to its broadcast nature, a single message is needed, 
regardless of the number of clients in the group.
However, reliable reception is guaranteed only for the relay client. 
The broadcast message sent by the GO is discarded at the IP layer by 
the legacy clients that are GOs of other groups; thus,  
it will not propagate outside the group where it has been generated. 
In order to advertise the content to other groups, the relay client sends
a content advertisement message to each legacy client that is also a GO
of another group, and waits for the ACK.
The example of Fig.~\ref{fig:proto} shows the sequence of application-layer
packets that are exchanged when new content becomes available at
Client 1A. After message \ding{199}, all nodes in the two groups 
have updated their own CRT with the new content item. 

Based on the above procedure, updated content information can be (surely) found
only at the GOs and at the relay clients. Thus, upon
generating a content request, a device that is neither a GO nor a
relay client, first looks up the content information in its CRT. If no
entry is available, it sends the request to its GO, which will process
it as described in the previous section.

\begin{journal}

\subsection{Message format}

\begin{figure}
	\centering
	\includegraphics[clip,width=7cm,angle=0]{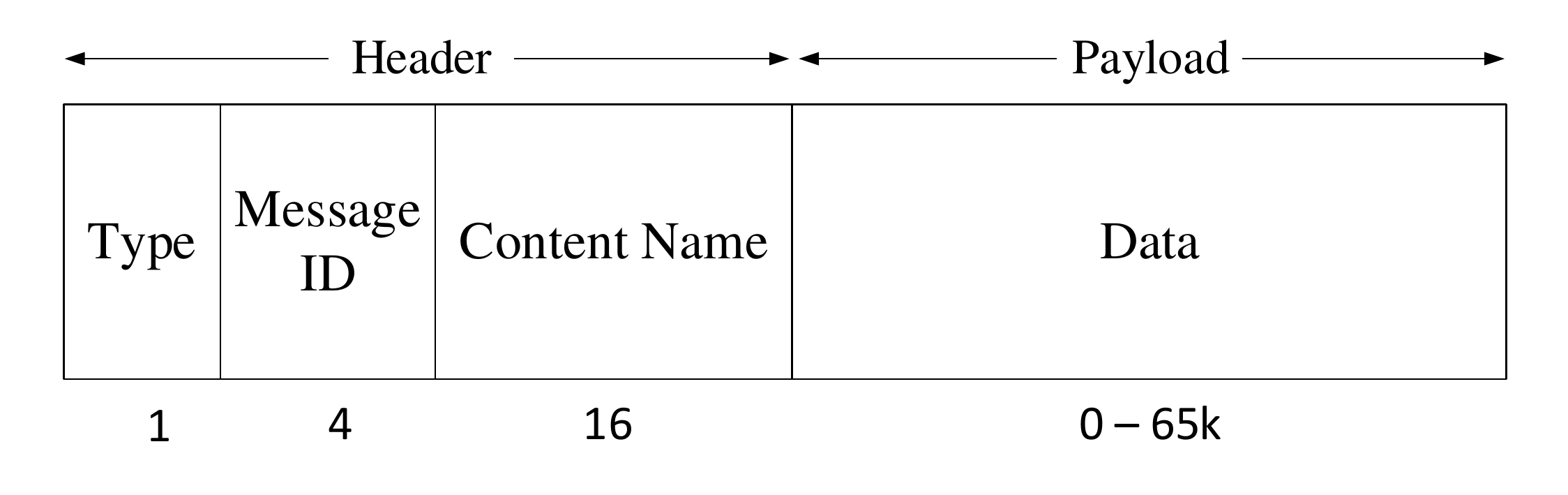}
	\caption{Application-layer message for content registration, advertisement, request and delivery. \label{fig:pfor}}
\end{figure}

Our content-centric routing is based on one application-layer message,
which may carry control information (e.g., content registration
and advertisement), a content request, or
the desired content item. The message is 
encapsulated in standard TCP or UDP segments, carried by IP packets.
Fig.~\ref{fig:pfor} reports the message format:  
\begin{itemize}
\item Type (1 byte) specifies the message type, among: Content
registration (and ACK), Content advertisement (and ACK), Content data, Content request,
Relay election (and ACK), notification of GO role in another group by a legacy client (and ACK). 
\item Message identifier (4 bytes) is a random nonce that associates the
ACK with the message it refers to.
\item Content name (16 bytes) is the MD5 hash of the content name. Note
that any other hash function could be used to encode the name.
\item Data (0-65 kbytes) is the main payload, carrying 
control or content data. 
\end{itemize}
\end{journal}

%% file: performance.tex
\section{Experimental Evaluation\label{sec:performance}}

We set up a testbed  including several Android devices of different type, namely,  Google Nexus 7 and ASUS Transformer Pad TF300
tablets  and 2 different smartphones (LG P700, Sony Xperia Miro ST23i). 
The Nexus tablets were equipped with Android 4.4.2 (API level 19), but our application was also tested with Android 4.3 (API level 18) on the same devices, before the operating system upgrade. 
LG smartphones used Android 4.0 (API level 14), which is the oldest version supporting Wi-Fi~Direct.
In our tests, LG smartphones acted as P2P clients and never as GOs, since the transport-layer tunnels from/to the GO discussed in Section~\ref{sec_log} are fully enabled only for Android 4.3 and later versions. 
The ASUS tablets and the Sony Xperia were equipped with Android 4.2.1 (API level 17) and Android 4.0.4 (API level 14), respectively. Neither of them support Wi-Fi~Direct; we used such devices only as legacy clients and not as group owners.   
This variety in the choice of  devices  allowed us to validate our 
multi-group communication mechanism in presence of heterogenous devices and different conditions. 
{\em No device was rooted}, to be sure that we could validate the approach for off-the-shelf devices.

We developed an Android application to implement our solution for bidirectional, multi-group communication and content-centric routing, as well as to validate the whole approach and  
assess its performance.
In order to program the devices, we used the integrated development environment
(IDE) Eclipse (version v22.0.1) with the ADT (Android Developer Tools)
on Ubuntu 13.04. The ADT is officially provided by Google and allows
users to build, test, and debug applications on Android.

\begin{figure}[!tb]
\centering
\includegraphics[clip=true,trim=1cm 3cm 1cm 3cm, width=5cm]{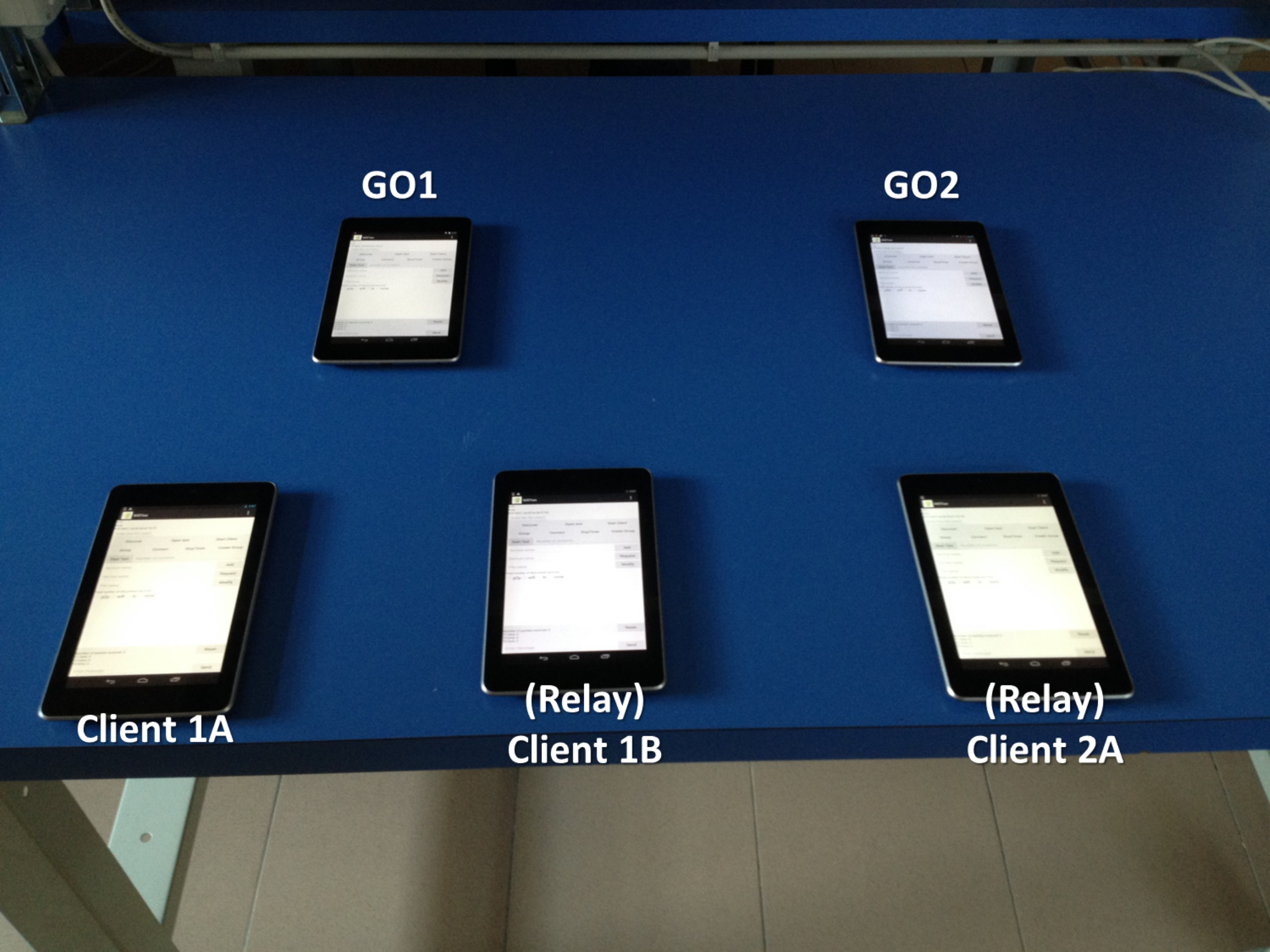}
\caption{Testbed setup with five tablet devices, forming two \wfd groups.\label{fig:testbed}}
\end{figure}

For brevity and ease of presentation, in the following we show the
results that we obtained using the experimental setup depicted
in Fig.~\ref{fig:testbed}, i.e., two \wfd groups.  Group 1  includes 4
devices (GO1, Client 1A, Client 1B and GO2, the latter acting as legacy
client in Group 1), while Group 2 comprises 2 devices (GO2 and Client
2A). Client 1B and Client 2A operate as relay clients in their own
groups.  This setup is equivalent to the scenario discussed in the
example of Fig.~\ref{fig:proto}. 

All the tablets were located in proximity of each other, to reduce the
effects of  propagation delays and signal attenuation due to distance.
All  experiments have been carried out in the laboratory, during evening
hours to reduce interference from active neighboring APs.  We manually chose
channel 11 for  \wf and \wfd communications, since, according
to a preliminary monitoring of the spectrum, it appeared to be the least
interfered channel. The interference power we measured on channel 11 was
always between -91~dBm and -98~dBm (i.e., quite low) during all the
experiments. As a term of comparison, the most crowded channels showed
levels of interference between -65~dBm and -83~dBm.

After the \wfd groups were set up, we tested the content-oriented
routing scheme in two phases. In the initial phase we investigated the
performance only of the content delivery scheme (i.e., of the forwarding
mechanism through transport-layer tunnels), while in the second phase we
 tested the content registration and advertisement mechanism.

\subsection{Content delivery performance}\label{sec:wfd_s}

Here, we focus on the performance that can be achieved for the content
data transfer, from one device to another, based on the data delivery
scheme explained in Section~\ref{sec:mechanism}. We manually configured 
the CRT and PIT tables to avoid any protocol overhead due to content
requests and table updating.  Each content is divided into chunks of
fixed size equal to 1400 bytes, to avoid IP fragmentation. To vary the
offered traffic load, the content provider periodically sends a new
chunk, encapsulated into a Content Data message, with the chunk rate
being a varying application parameter. 

We validated the data delivery mechanism by picking different pairs of
devices among the possible ones, and letting them act as
source-destination nodes. We therefore verified the full bidirectional
connectivity over the whole multi-group network, and recorded the
application-layer throughput and the packet losses experienced at the IP
layer,  as functions of the application-layer traffic offered load. 
For each configuration, we run 100 different experiments, to obtain throughput results with a 1\% relative width of the 95\% confidence interval. 

In the following, we mainly focus on the scenarios detailed below, always run on the testbed 
shown in Fig.~\ref{fig:testbed}:
\begin{enumerate}
\item ``2 devices - 1 group'' (2d1g), in which the source is Client 1A and the destination is GO1. The communication between a client and its GO involves just one hop at IP and MAC layer, since  each message is sent through a single unicast IP packet, carried by a single MAC frame.   
\item ``3 devices - 1 group'' (3d1g), in which the source is Client 1A and the destination is Client 1B. The communication between two clients in the same group involves one hop at the IP layer, but two hops at the MAC layer (Client 1A $\to$ GO1 $\to$ Client 1B).
\item ``4 devices - 2 groups'' (4d2g), in which the source is Client 2A and the destination is Client 1B. The communication between the two clients in 2 groups requires two hops at IP layer (Client 2A $\to$ GO2 $\to$ Client1B) and three hops at MAC layer (Client 2A $\to$ GO2 $\to$ GO1 $\to$ Client 1B).
\item ``2 devices - 1 group - broadcast'' (2d1g-B), in which the source is GO2 and the destination is Client 2A. The communication within the same group now occurs in the opposite direction with respect to the 2d1g case, but notably the single-hop communication is based on a broadcast transmission, since GO2 is also legacy client of GO1. 
\item ``4 devices - 2 groups - broadcast" (4d2g-B), in which the source is Client 1B and the destination is Client 2A. The communication between the two clients in two different groups involves 2 hops at IP layer (Client 1B $\to$ GO2 $\to$ Client 2A) and 3 hops at MAC layer (Client 1B $\to$ GO1 $\to$ GO2 $\to$ Client 2A) in which the last hop is based on a broadcast transmission.
\end{enumerate}
Note that we do not show the case of content transfer from GO1 to Client
1A since it is equivalent to the 2d1g case; indeed, GO1 is not a legacy
client of any other group, thus it can send unicast IP packets directly
to Client 1A.

For fair comparison,  we start by evaluating the first three cases,
which imply only unicast transmissions; then, we will move on to the last
two cases, involving broadcast transmissions. 


\begin{figure}[!tb]
\centering
\includegraphics[width=0.85\linewidth]{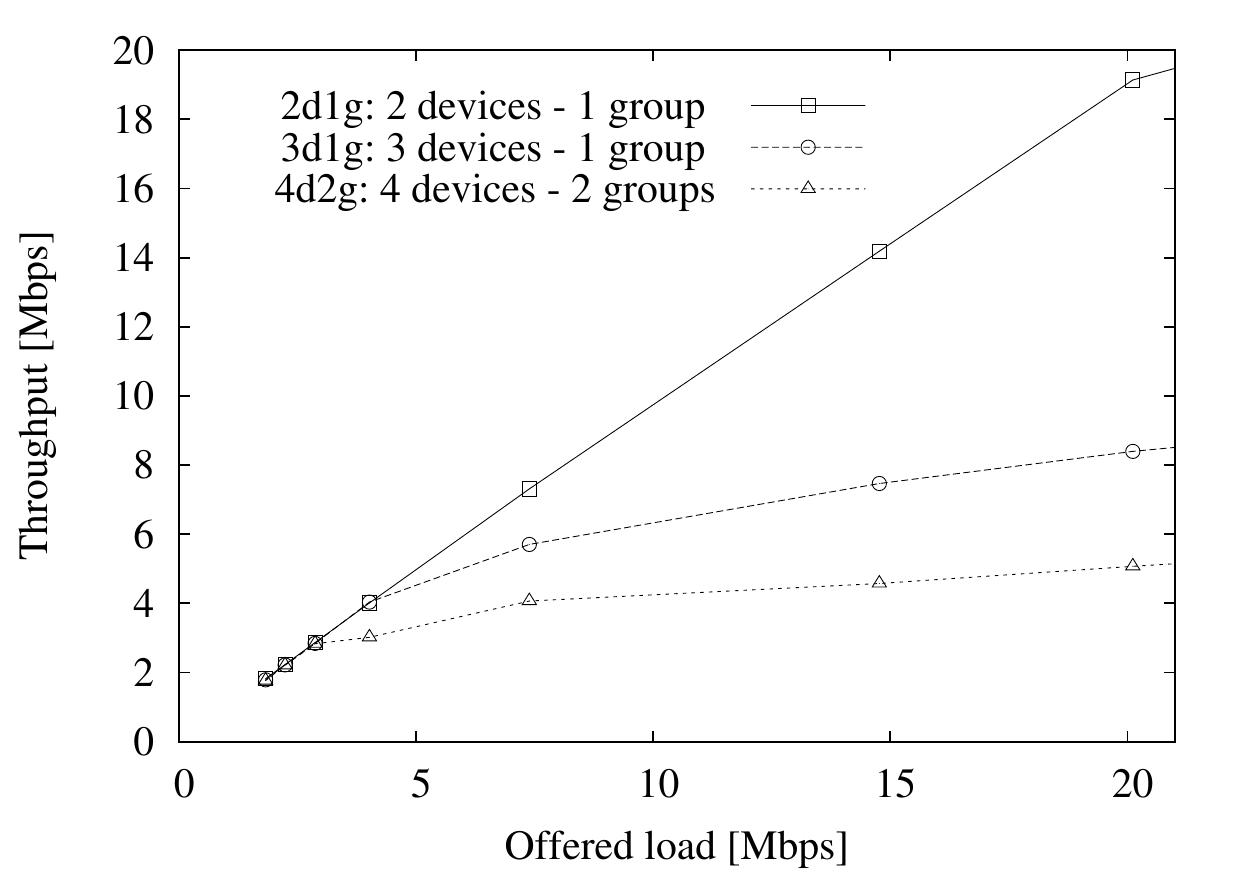}
\caption{Throughput at application layer as a function of the offered traffic load, for packet  transfers involving only unicast transmissions.\label{fig:th2dev}}
\end{figure}

\begin{journal}
\begin{figure}[!tb]
\centering
\includegraphics[scale=1,width=1\linewidth]{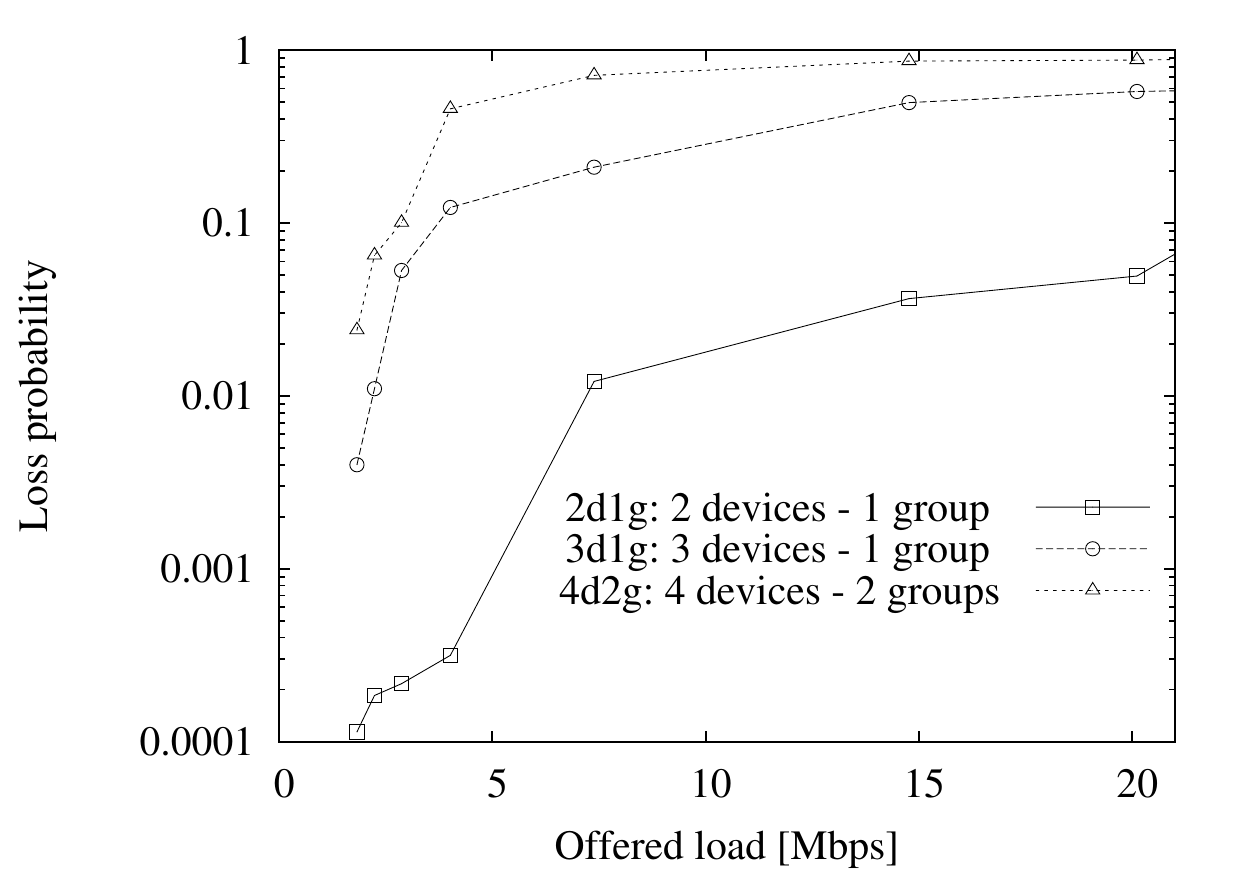}
\caption{Packet loss at IP layer vs.\ offered traffic load, for packet  transfers involving only unicast transmissions.\label{fig:pl2dev}}
\end{figure}
\end{journal}

\begin{figure}[!tb]
\centering
\includegraphics[width=0.85\linewidth]{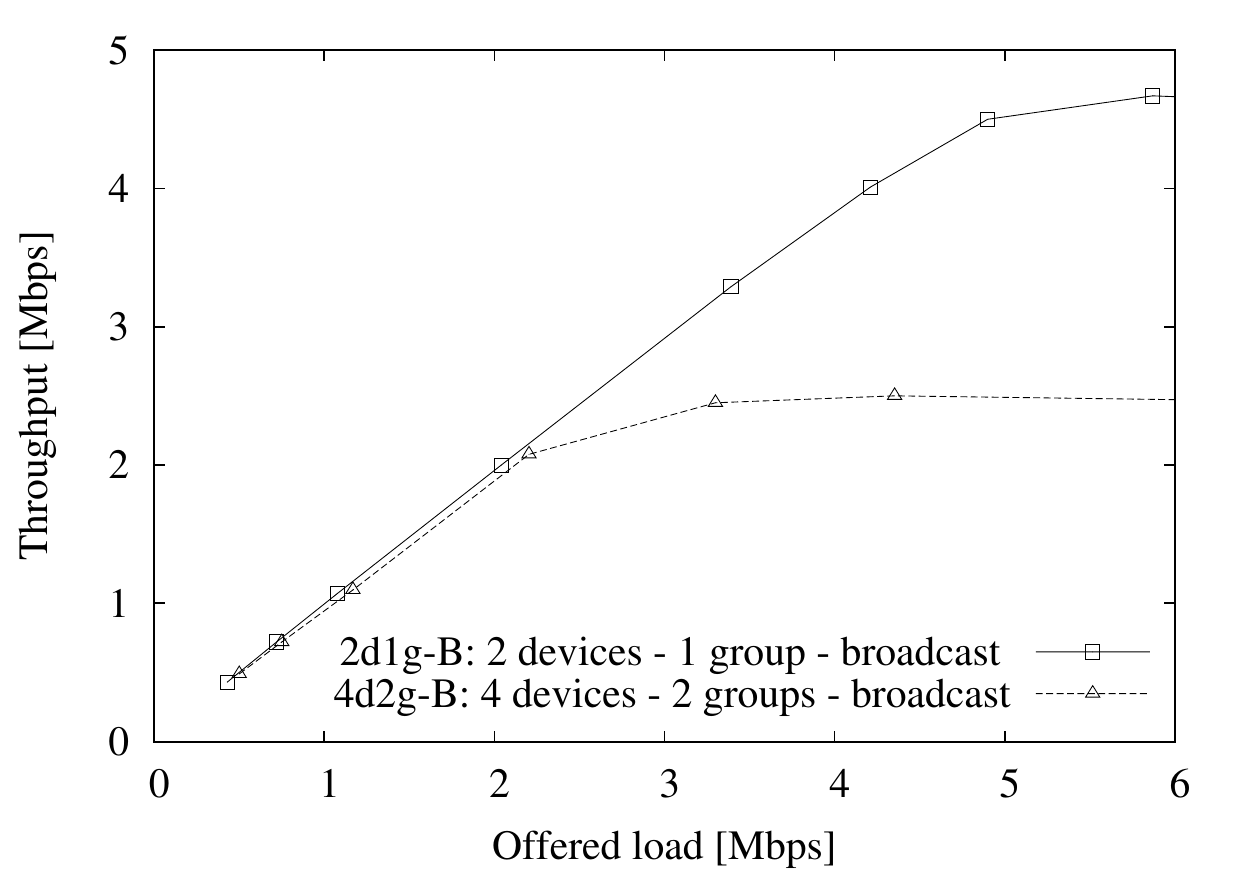}
\caption{Throughput at application layer as a function of the offered traffic load, for packet  transfers involving also broadcast transmissions.\label{fig:th-back}}
\end{figure}

\begin{journal}
\begin{figure}[!tb]
\centering
\includegraphics[scale=1,width=1\linewidth]{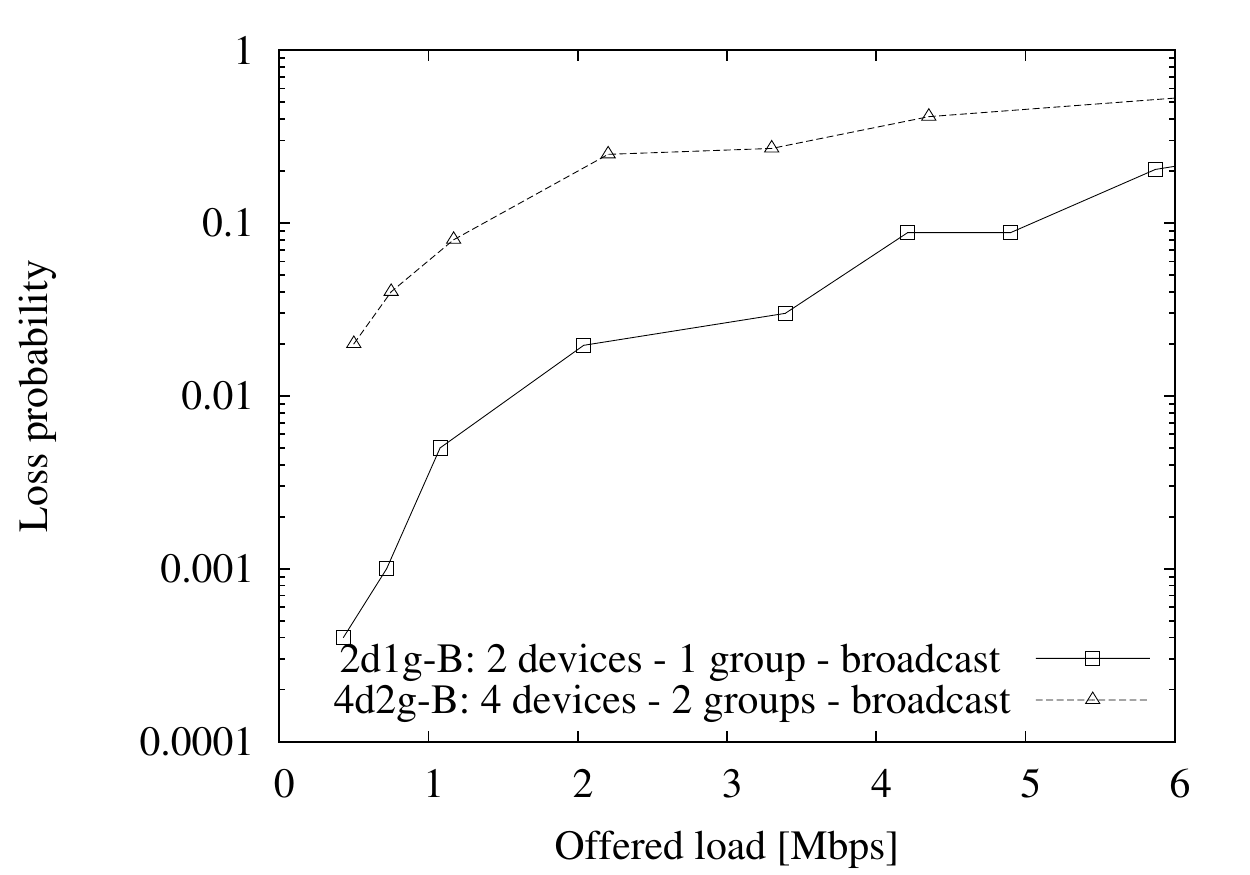}
\caption{Packet loss at IP layer vs.\ offered traffic load, for packet  transfers involving also broadcast transmissions.\label{fig:loss-back}}
\end{figure}
\end{journal}

Fig.~\ref{fig:th2dev} shows the application-layer throughput vs. the
offered load. As expected, the throughput increases with the load, and
reaches a maximum value of about 19~Mbit/s (2d1g scenario), 8.4~Mbit/s
(3d1g scenario) and 5.0~Mbit/s (4d2g). These results are coherent with
the fact that the throughput decreases proportionally to the number of
hops, due to the channel contention among the transmitters operating
on different hops. Note that current available \wfd interfaces work only
on a single frequency channel and, thus,  the whole multi-group network
is part of the same collision domain. In general, the number of hops
traversed by a packet depends only on the distance, in terms of number
of groups, over the backbone between source and destination (usually, we
have two hops at the MAC layer per each traversed group), whereas it is
independent of the total number of devices composing  the network. 
While the single collision
domain increasingly affects the performance as the number of active
transmitters grows, having the hop number independent of the group size
improves scalability. Additionally, the impact of the 
single collision domain lessens as the network gets larger: 
when  transmitters are far away from each other, 
some degree of spatial diversity is possible and
interference among  parallel  transmissions greatly reduces. It
follows that  the network throughput decreases less than
proportionally to the number of hops.

\begin{journal}
Fig.~\ref{fig:pl2dev} shows the overall packet loss probability. Note
that  packet losses are almost negligible  in the 2d1g scenario. They
become noticeable for the 3d1g case ($0.4\%$ for the lowest load) but
still have little impact on the throughput. Under the 4d2g scenario,
instead, packet losses are more significant ($2.4\%$ for the lowest
load), since the transmissions over the three hops that every message
has to undergo at the MAC layer interfere with each other.
\end{journal}

\begin{journal}
Figs.~\ref{fig:th-back} and~\ref{fig:loss-back} depict throughput and
packet losses, respectively, 
\end{journal}
\begin{conference}
{\color{red}
Fig.~\ref{fig:th-back} depicts the throughput 
}
\end{conference}
for the last two scenarios, 2d1g-B and
4d2g-B, both implying one broadcast transmission by GO2. The maximum
throughput is 4.6~Mbit/s for 2d1d-B and 2.5~Mbit/s for 4d2g-B. Such
numbers are much smaller than in the first three scenarios, since, at MAC
layer, 802.11 broadcast packets are transmitted  at the minimum data rate
 (6 Mbit/s), whereas  much higher rates are used for unicast
transmissions (up to 54~Mbit/s). Note also that, even if three hops are
involved in 4d2g-B, the first two hops occur through unicast
transmissions (hence at much higher data rate than broadcast
transmissions) and, thus, they mildly  affect the throughput. 
\begin{journal}
Looking at
Fig.~\ref{fig:loss-back}, it can be seen that the loss probability is
higher in the two scenarios with broadcast transmissions than in
previous cases. This behavior is also expected:  in case of failure,
broadcast packets are never retransmitted at the MAC layer, thus the
reliability of the communication from the GO to its relay client is
severely reduced.
\end{journal}

In summary, the performance of the communication backbone is strongly
affected by the traffic flow direction. The two different relay schemes,
adopted within a group to work around the constraints imposed by
Wi-Fi~Direct, show significantly different performance. The main
bottleneck is represented by broadcast communications from the GOs to
their relay clients.

\subsection{Content registration and advertisement performance}

We now investigate the performance of the process for content
registration and advertisement, by programming one device to
periodically register a new content item and measuring the latency
experienced at each node to update its own CRT.

We focus on the scenario in Fig.~\ref{fig:proto} where, every second,
Client 1A registers one new content item with GO1, through the two
messages \ding{192}, \ding{193} reported in the figure. The experiment
lasted 1 minute, with a total of 60 new registered items. The sequence
of the advertisement messages that are generated and transmitted is
represented by  messages \ding{194}-\ding{199}; all of them are
processed at the application layer. Table~\ref{tab:adv} reports the
latency measured at each hop, as well as the  end-to-end latency in the
4d2g scenario, obtained by logging the time at which each device
processes the incoming advertisement message. Such message triggers a CRT
update and the transmission of the corresponding ACK. Clock offsets 
affecting the logs of
different devices were computed through a packet-level trace
obtained by an external laptop sniffing traffic in monitor mode.


\begin{table}[!tb]
\caption{Latency to advertise a new content item in the two-group scenario of  Fig.~\ref{fig:proto}}\label{tab:adv}
\begin{center}
\begin{tabular}{|c|c|c|c|}
\hline
Transfer & Incoming & Average & 95\% confidence\\
         &  message & [ms] &  interval [ms]\\
\hline
GO1 $\to$ Client 1B & \ding{194} & 250  &   206-294\\
Client 1B $\to$ GO2  & \ding{196} & 304  &   219-390\\
GO2 $\to$ Client 2A  & \ding{198} & 226  &   199-252 \\
\hline
GO1 $\to$ Client 2A  & \ding{194} & 780 &  688-872 \\
\hline
\end{tabular}
\end{center}
\end{table}


The overall latency required to update the farthest node 
is less than one second. According to our data, the main contribution is
due to the processing time at the application layer of each node since
the transmission time of the advertisement messages (and their ACKs) is
negligible. This points to the need for optimized versions of relay code
running at backbone nodes.


\subsection{Comparison with Bluetooth}
{

The Android devices we used in the previous experimentation are equipped
with Bluetooth~3.0, which also supports multi-hop communications. For
this reason, we have chosen to compare the performance of multi-group
communications in \wfd and in Bluetooth under similar scenarios.

Under the same testbed setup shown in Fig.~\ref{fig:testbed}, we run a
logical topology in Bluetooth, shown in Fig.~\ref{fig:topo-bt}, to mimic
exactly the \wfd multi-group topology, considered in the previous
sections. We set up two piconets, P1 and P2. P1 consists of three
devices: one master (M1) and two associated slaves (S1A, S1B). P2
consists of two devices: one Master (M2) and one associated Slave (S2A).
To enable bridging capabilities among the two piconets, M2 is also
connected to M1 as a slave. We developed a single Android application
({\em not requiring to root the devices}) that can generate traffic,
relay packets between the two piconets and record performance
metrics. We run this application on each device, manually configuring
the role of each device (source/destination/gateway).

We focused mainly on the maximum achievable throughput in different
scenarios, when changing the source-destination pairs. We adopted
990~bytes as the application-layer packet size, to avoid packet
fragmentation. The throughput is always measured at the receiver's
application layer. 


\begin{figure}[!tb]
\centering
\includegraphics[scale=1,width=0.6\linewidth]{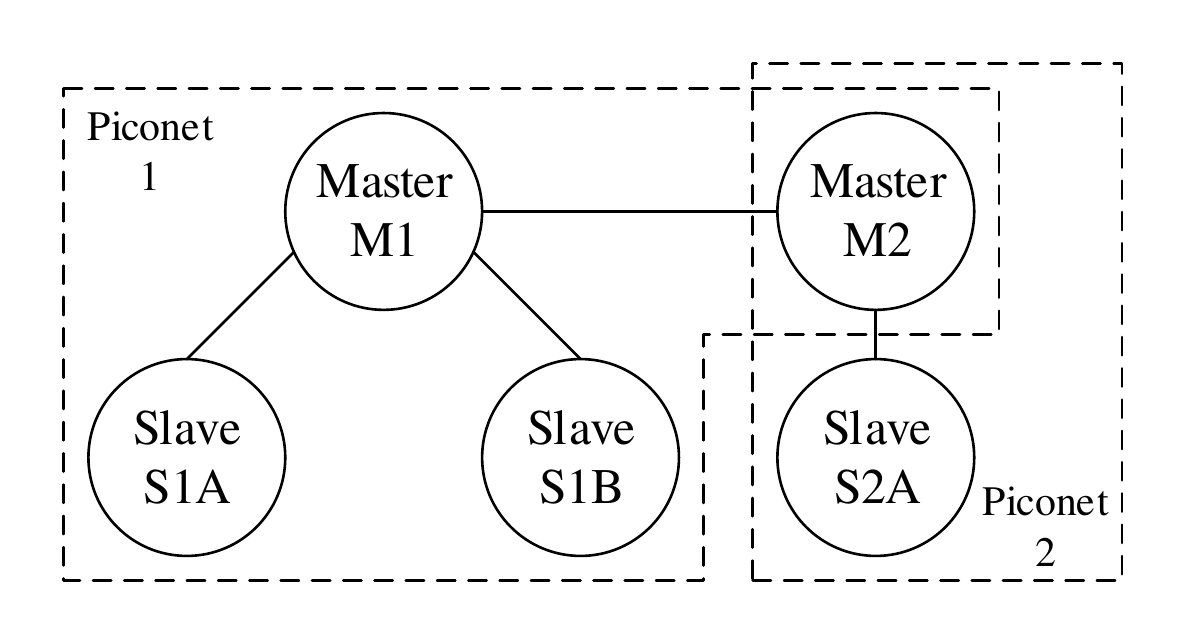}
\caption{Logical topology for the scenario based on Bluetooth communications, comprising two piconets. \label{fig:topo-bt}}
\end{figure}

To faithfully mimic the scenarios considered in Sec.~\ref{sec:wfd_s}, we
define each piconet as a ``group'' and consider the following cases:
\begin{enumerate}
\item ``2 devices - 1 group'' (2d1g), in which the source is S1A and the
destination is M1. The packets are directly sent from S1A to M1 at the MAC
layer without any relay.
\item ``3 devices - 1 group'' (3d1g), in which the source is S1A and the
destination is S1B. The traffic generated by S1A is first sent to the
M1, i.e.\, the piconet master. Each packet is processed by M1 at
application layer and then relayed to S1B. The overall communication
involves two hops at both application and MAC layer (S1A $\to$ M1 $\to$
S1B). 
\item ``4 devices - 2 groups'' (4d2g), in which the source is S2A and
the destination is S1B. The traffic traverses P1 and P2 thanks to the
two application-layer relays operated by the two masters, M2 and M1.
The overall communication involves 3 hops both at application and MAC
layer (S2A $\to$ M2 $\to$ M1 $\to$ S1B). 
\end{enumerate} 
Note that the Bluetooth does not support broadcast, hence we do not
consider the last two scenarios in Sec.~\ref{sec:wfd_s} involving
broadcast communications.

We now compare the maximum throughput between \wfd and
Bluetooth. Table~\ref{tab:bt-wfd} provides the maximum throughput
achieved in each scenario, measured at application layer. The throughput
is expressed in terms of absolute value and normalized value, as
described below.

For Bluetooth, the maximum absolute throughput with 2 devices in direct
communication (2d1g) is around 1.9~Mbit/s, which is consistent with the
maximum net data rate of 2.1~Mbit/s. For 2-hop communications
(3d1g), the throughput decreases by a factor of two (0.9~Mbit/s); this is
expected, since the communication slots used by master M1 are divided in
two: one to receive the data from one slave (S1A) and one to send the
data to the other slave (S1B). In the case of two piconets (4d2g), the
maximum throughput is slightly less than 3d1g, since simultaneous
transmissions (S1A$\to$M1 and M2$\to$S2A) can occur in the two different
piconets.

Table~\ref{tab:bt-wfd} reports also the throughput achievable by Wi-Fi\
Direct, which is much higher with respect to Bluetooth in absolute
terms, thanks to the higher data rates.
\begin{table}[!tb]
\caption{Experimental maximum throughput of \wfd and Bluetooth, measured at application layer\label{tab:bt-wfd}}
\centering
\begin{tabular}{| c || c | c | c ||c|c|c|}
  \hline
  & \multicolumn{3}{|c||}{Throughput Mbit/s} &
  \multicolumn{3}{|c|}{Normalized throughput} \\
Technology  & 2d1g & 3d1g & 4d2g & 2d1g & 3d1g & 4d2g \\
 & 1 hop & 2 hops & 3 hops & 1 hop & 2 hops & 3 hops\\
  \hline
  Bluetooth & 1.92 & 0.94 & 0.77 & 64\% & 31\% & 26\%\\
  \wfd & 22.9 & 10.6 & 6.2 & 35\% & 16\% & 9.5\%\\
  \hline
\end{tabular}
\end{table}
For a fair comparison, we also report the normalized throughput obtained
by dividing the throughput by the actual physical data rate adopted
during the communication. In \wfd the rate must adapt to the channel
conditions, and in this case we observed, most of the times, packets
sent at the maximum data rate (65~Mbit/s) thanks to the small physical
distance (always less than 40~cm) between the devices. For Bluetooth,
the physical data rate adopted for the normalization is 3~Mbit/s, which
is the data rate for  Bluetooth~3.0 operating in the devices. 

By considering only the normalized throughput in Table~\ref{tab:bt-wfd},
 \wfd achieves about 35\%, 16\% and 9.5\% of the maximum throughput,
respectively for increasing number of transmission hops, while Bluetooth
achieves about 64\%, 31\% and 26\% of it. The lower efficiency of \wfd
is due to the contention-based protocol that regulates the access to the
same radio channel. As already observed, the throughput decreases
proportionally to the number of transmission hops. Instead, in Bluetooth
the efficiency is larger thanks to the slotted time version of the
protocol, whose access is coordinated by the Master.
	}

%% file: relatedwork.tex
\section{Related Work\label{sec:relatedwork}}

Several recent studies have 
investigated the features and the performance  of the Wi-Fi Direct technology. 

One of the first studies has appeared in \cite{camps-mur2011}, where
Camps Mur et al. consider a single-group Wi-Fi Direct network with the
group owner sharing access to a 3G network with a set of connected
devices. The work analyzes  the power saving protocols defined in Wi-Fi
Direct and design two algorithms that use such protocols to save energy
while  providing good throughput performance. An improved power
management scheme for Wi-Fi Direct is proposed in \cite{Lim2013}, which 
dynamically adapts the duty cycle of P2P devices to the properties of
the application to be supported.

An overview and experimental evaluation of Wi-Fi Direct using two
laptops running Linux is presented in~\cite{camps-mur}, where the
emphasis is  on the standard group formation procedures and the
performance that they exhibit in terms of delay and power consumption.
Group formation is also the focus of the work in  
\cite{Conti2013}, which investigates the ability to create  
opportunistic networks of devices using Wi-Fi Direct to establish communication links. 
The performance of group formation is studied experimentally, by
varying the protocol parameters and considering scenarios that are typical of opportunistic
networks. A preliminary study of multi-group physical topologies of
Wi-Fi Direct networks can be found in our previous work \cite{wtc14},
where however only some  of the limitations of the Android OS are
investigated and only unidirectional D2D communication is tackled.

The use of Wi-Fi Direct as a D2D technology to be integrated 
into LTE and LTE-A cellular networks is explored in~\cite{Mancuso,Sergey1,Sergey2}. 
In particular, while \cite{Mancuso} mainly focuses on architectural
issues, \cite{Sergey1} and \cite{Sergey2} also quantify the estimated
network performance gains from offloading cellular traffic onto Wi-Fi
Direct-based, D2D connections.

As for content dissemination and sharing in mobile ad hoc networks, 
a number of solutions have been proposed in the literature, e.g., \cite{Eureka,Meroni,Tiago}. 
However, very few works exist that specifically address Wi-Fi Direct-based networks.
Among these, the study in~\cite{duong} presents a Wi-Fi Direct-based
overlay architecture 
for content sharing among peers belonging to the same group.
In particular, they leverage the P2PSIP protocol, which enables
real-time communication using the application-layer signaling protocol
SIP in a peer-to-peer fashion. The work in \cite{lombera}, instead,
implements the decentralized iTrust mechanism \cite{iTrust} for
information publication and retrieval. In particular, it proposes a peer
management technique to facilitate group creation and allow peers to set
up and maintain  connectivity over Wi-Fi Direct. 

As mentioned, to the best of our knowledge, none of the existing works
has investigated, solved and experimentally evaluated  bidirectional
communication in Wi-Fi Direct multi-group networks.

%% file: conclusion.tex

\section{Conclusions and Future Work\label{sec:conclusions}}

We implemented bidirectional, multi-group communication in Android devices supporting the recent \wfd protocol. 
This allowed us to extend the achievable communication range for a protocol whose current implementation 
in off-the-shelf, unrooted Android devices has been tailored just to single group D2D communication.

In particular, we proposed a solution to overcome the limitations of the
physical \wfd network topology and of its addressing plan, and we built
a logical topology that enables bidirectional inter-group data
transfers. The logical topology we devised is based on a cooperative
traffic relaying scheme among adjacent groups and, through
transport-layer tunnels, leads to the formation of a network backbone
that provides full network connectivity. We also devised a
content-centric routing scheme, which properly exploits the above
backbone and allows content advertisement, discovery and retrieval in
arbitrary D2D network topologies. We implemented our solution in Android
and validated it by developing a testbed comprising a heterogenous set
of devices. 


Our work opens up several future research directions. Firstly, an
in-depth study could be carried out to determine the system scalability
with the number of network devices. Such study could also factor in the
choice of nodes  to be elected as relay clients (their number and
typology) as well as the techniques to efficiently manage the
consequences of node churning. Secondly,  our data transfer mechanism
can support distributed strategies for the formation of smart topologies
involving multiple groups and covering extended geographical areas.
Lastly, bidirectional, inter-group communication can be the basis for
disruptive cooperative applications and service models.

\section{Acknowledgements}

This work was partially supported by Telecom Italia S.p.A. under the ``Smart Connectivity'' Research Contract.